\documentclass[fleqn,usenatbib]{mnras}

\usepackage{newtxtext,newtxmath}

\usepackage[T1]{fontenc}

\DeclareRobustCommand{\VAN}[3]{#2}
\let\VANthebibliography\thebibliography
\def\thebibliography{\DeclareRobustCommand{\VAN}[3]{##3}\VANthebibliography}

\usepackage{graphicx}	
\usepackage{amsmath}	
\usepackage[euler]{textgreek}
\usepackage{tablefootnote}

\title[Digging deeper into NGC\,6868]{Digging deeper into NGC\,6868 II: ionized gas and excitation mechanism}

\author[J. P. V. Benedetti et al.]{
João P. V. Benedetti,$^{1,2,3}$ \thanks{E-mail: jbenedetti@iac.es(JPVB)}
Rogério Riffel,$^{1}$ \thanks{E-mail: riffel@ufrgs.br (RR)}
Tiago Ricci,$^{4}$
Rogemar A. Riffel,$^{5}$
Miriani Pastoriza,$^{1}$
\newauthor
Marina Trevisan,$^{1}$
Luis G. Dahmer-Hahn,$^{6}$
Daniel Ruschel-Dutra,$^{7}$
Alberto Rodríguez-Ardila,$^{8}$ \newauthor
Anna Ferr\'e-Mateu,$^{2}$
Alexandre Vazdekis$^{2,3}$ and
João Steiner$^{9}$ \thanks{In Memorian.}
\\
$^{1}$Departamento de Astronomia, Universidade Federal do Rio Grande do Sul. Av. Bento Gonçalves 9500, 91501-970 Porto Alegre, RS, Brazil\\
$^{2}$Instituto de Astrof\'\i sica de Canarias, Calle V\'\i a L\'actea s/n, E-38205 La Laguna, Tenerife, Spain\\
$^{3}$ Departamento de  Astrof\'\i sica, Universidad de La Laguna, E-38205, Tenerife, Spain \\
$^{4}$Universidade Federal da Fronteira Sul, 97900-000 Campus Cerro Largo, RS, Brazil\\
$^{5}$Departamento de Física, Universidade Federal de Santa Maria, Centro de Ciências Naturais e Exatas, 97105-900 Santa Maria, RS, Brazil\\
$^{6}$Shanghai Astronomical Observatory, Chinese Academy of Sciences, 80 Nandan road, Shanghai 200030, China\\
$^{7}$Departamento de Física - CFM - Universidade Federal de Santa Catarina, 476, 88040-900 Florianópolis, SC, Brazil\\
$^{8}$Laboratório Nacional de Astrofísica/MCT - Rua dos Estados Unidos 154, Bairro das Nacões. CEP 37504-364 Itajubá, MG, Brazil \\
$^{9}$Instituto de Astronomia, Geofísica e Ciências Atmosféricas, Universidade de São Paulo, 05508-900 São Paulo, Brazil 
}

\date{Accepted XXX. Received YYY; in original form ZZZ}

\pubyear{2024}

\begin{document}
\label{firstpage}
\pagerange{\pageref{firstpage}--\pageref{lastpage}}
\maketitle

\begin{abstract}
 
We studied the ionized gas in the inner region (\(\sim\)$680\times470$~{pc\textsuperscript{2}}) of the galaxy NGC\,6868 using Gemini/GMOS integral field unit observations. Channel maps reveal complex kinematics and morphology, indicating multiple processes at work in NGC\,6868. Through emission-line fitting, we identified two ubiquitous components in our data: a narrow ($\sigma\sim110$~{km\:s$^{-1}$}) tracing an ionized gas disc and a broad component ($\sigma\sim300$~{km\:s$^{-1}$}) mainly associated with inflowing/outflowing gas. The derived V-band reddening shows a spatial distribution consistent with that obtained from stellar population synthesis, although with generally higher values. For the first time, we measured the electron temperature in NGC\,6868, finding values ranging from \(\sim\)$14000$~{K} in the central region to $\gtrsim20000$~{K} with an outward increasing temperature gradient. The electron density map exhibits an inverse relationship, with central values reaching $N_e\sim4000$~{cm\textsuperscript{-3}} for the broad component decreasing to $N_e\sim100$~{cm\textsuperscript{-3}} towards the edges of the field of view. Using BPT diagrams, we found that all spaxels are consistent with both AGN and shock ionization. However, when this information is combined with our kinematic and temperature findings, and further supported by the WHAN diagram, we argue that an AGN is the dominant ionisation mechanism in the central region of NGC 6868, while the extended outer component is ionized by a combination of hot low-mass evolved stars and shocks. According to our findings, shocks play a significant role in the ionization balance of this galaxy.
\end{abstract}




\section{Introduction}
    \label{sec:intro}

    Since the discovery of tight correlations of the central black-hole (BH) mass with galaxy properties, such as stellar velocity dispersion and bulge mass \citep[][]{MagorrianEtAl1998, GebhardtEtAl2000, HaringRix2004} and with the ever-growing evidence that the different manifestations of active galactic nuclei (AGN) were the counterpart of BHs, a picture of co-evolution and interaction between the two components has emerged \citep[][]{Fabian2012, KormendyHo2013, HeckmanBest2014}. The injection of energy by the most energetic AGN in the interstellar medium (ISM) plays a crucial role in quenching the star formation (SF) mainly in the most massive galaxies in cosmological simulations \citep[e.g.][]{CrotonEtAl2006, SegersEtAl2016} by heating and expelling the available gas for star-formation. However, observational studies trying to link SF and AGN activity are still inconclusive.
    
    The majority of studies trying to establish this link focused on relatively bright objects \citep[e.g. Seyferts and quasars, e.g.][]{NayakshinZubovas2012} and the effects of low-luminosity AGN (LLAGN) in the circumnuclear stellar population is even less studied. Many LLAGNs are classified as LINERs (Low-Ionization Nuclear Emission Regions) and proved to be numerous in the nearby Universe, being present in 1/3 of all galaxies and corresponding to 2/3 of all different types of AGN \citep[][]{Ho2008}. Hence detailed studies of the circumnuclear region are needed to improve our knowledge of the impact of these sources in their vicinity.
    
    LINERs were formerly described by \citet{Heckman1980} as having strong lines from low-ionization species (e.g. \ion{O}{i}) and weaker high ionization lines (e.g. \ion{O}{iii}). Over the decades with compiling evidence \citep[][]{FerlandNetzer1983, HalpernSteiner1983, HoEtAl1996, HoEtAl1997, ConstantinVogeley2006} a growing consensus was formed over the picture that they were toned down versions of Seyfert nuclei, meaning the ionized gas features were due to the photoionization of an LLAGN. Detection of parsec-scale radio nuclei in 50\% of LINERs and subparsec jets \citep[][]{NagarEtAl2005} and, more recently, the detection of ionized gas outflows in approximately \(\sim\)46\% of LINERs in the sample from \citet[][]{HermosaMunozEtAl2022} related with the central supermassive black-hole further endorsed this picture. On the other hand, X-ray studies hint that AGN could not be solely responsible for the optical emission-line intensities observed \citep[][]{FlohicEtAl2006}.
    
    With improved spatial resolution studies, LINER-like signatures were found not only in the nuclear regions of galaxies ($<1$~{kpc}) but also at greater distances, hence the terms LIER or LI(N)ER adopted by some authors \citep[e.g.][]{SinghEtAl2013, BelfioreEtAl2016}. In this scenario, other ionization mechanisms need to be taken into account, such as shocks \citep[galactic or from an outflow,][]{Heckman1980, DopitaSutherland1995, HoEtAl2014, HoEtAl2016}, starbursts dominated by Wolf-Rayet stars \citep[][]{BarthShields2000} and post-asymptotic giant branch stars \citep[pAGB,][]{BinetteEtAl1994}. The latter scenario has been gaining support as the dominant ionization mechanism in objects with LINER-like extended emission \citep[][]{StasinskaEtAl2008, CidFernandesEtAl2011, YanBlanton2012, SinghEtAl2013}. Compelling evidence has been found in red-and-dead galaxies \citep[][]{HsiehEtAl2017}, such as a correlation between the H$\alpha$ surface density and the stellar population. This hints at a stellar origin for the LINER signature and the lack of ionizing photons due to LLAGN to explain the LINER spectra \citep[][]{EracleousEtAl2010}. In this sense, LINERs (or LIERs) can be found in different objects and phenomena in different parts of the spectra and, despite the similarity in spectral signatures, cannot be grouped as a homogeneous class \citep[][]{HerpichEtAl2016}.
    
    Integral Field Spectroscopy (IFS) has been used in the past decade to improve our understanding of these objects as they are a powerful tool to disentangle the different ionization mechanisms present in objects with LI(N)ER-like emission. \citet{SarziEtAl2010}, using the SAURON (Spectroscopic Areal Unit for Research on Optical Nebulae) IFS survey, 
    found a tight correlation between the stellar mass surface density and $\text{H}\beta$ surface density hinting at a stellar origin behind the ionization process. \citet[][]{LoubserSoechting2013} found that LLAGN can explain the observed ionization, but shocks and pAGB photoionization could not be ruled out. \citet[][]{RicciEtAl2014, RicciEtAl2014a, RicciEtAl2015, RicciEtAl2015a} extensively analysed a group of 10 LI(N)ER galaxies from a range of morphological types. In these objects, they found convincing AGN presence in at least 8 of the objects. However, they also report discs of ionized gas in 7 of them, with 3 having pAGB stars as the most probable source of ionizing photons for this disc component. \citet[][]{BelfioreEtAl2015} studied 14 galaxies and found extended ionized gas components, consistent with emission from pAGB stars. \citet[][]{HsiehEtAl2017} using a larger subset of MaNGA (Mapping Nearby Galaxies at APO) also found a correlation between the stellar surface density and H$\alpha$ surface density, indicating the same scenario as the one derived using SAURON. Studies using CALIFA data \citep[Calar Alto Legacy Integral Field Area Survey, ][]{KehrigEtAl2012, PapaderosEtAl2013, SinghEtAl2013, GomesEtAl2016} found "ubiquitous hot evolved stars and rare accreting black-holes". \citet[][]{LagosEtAl2022} using MUSE data to study group-dominant early-type galaxies found that the central regions of these objects are more influenced by LLAGNs with outer regions ionized by pAGB stars. A more recent study from \citet[][]{MenezesEtAl2022} using the mini-DIVING\textsuperscript{3D} (Deep IFS View of Nuclei of Galaxies) sample was able to separate the nuclear from the circumnuclear region, thus allowing a different treatment for each component due to the high spatial resolution for their data. From their sample, 23\% present LINER-like emission of which 69\% have signs of AGN activity. \citet[][]{RicciEtAl2023} analysed the nuclear region of 56 early-type galaxies contained in the DIVING\textsuperscript{3D} and classified $52\pm7$~{\%} of them into LINER/Seyfert, detecting broad components in H$\alpha$ in at least $29\pm7$~{\%}. With the addition of multi-wavelength data from other works \citep[][]{RampazzoEtAl2013, SheEtAl2017, BiEtAl2020}, out of the 48 ETGs with emission lines detected, 41 have compelling evidence of an AGN. One remarkable result is that in their sample, they found no transition objects, which are thought to encompass objects with LINER/\ion{H}{ii} mixture \citep[][]{HoEtAl1993, KewleyEtAl2006}. 

\begin{figure*}
    \centering
    \includegraphics[width=1.0\textwidth,height=1.0\textheight, keepaspectratio]{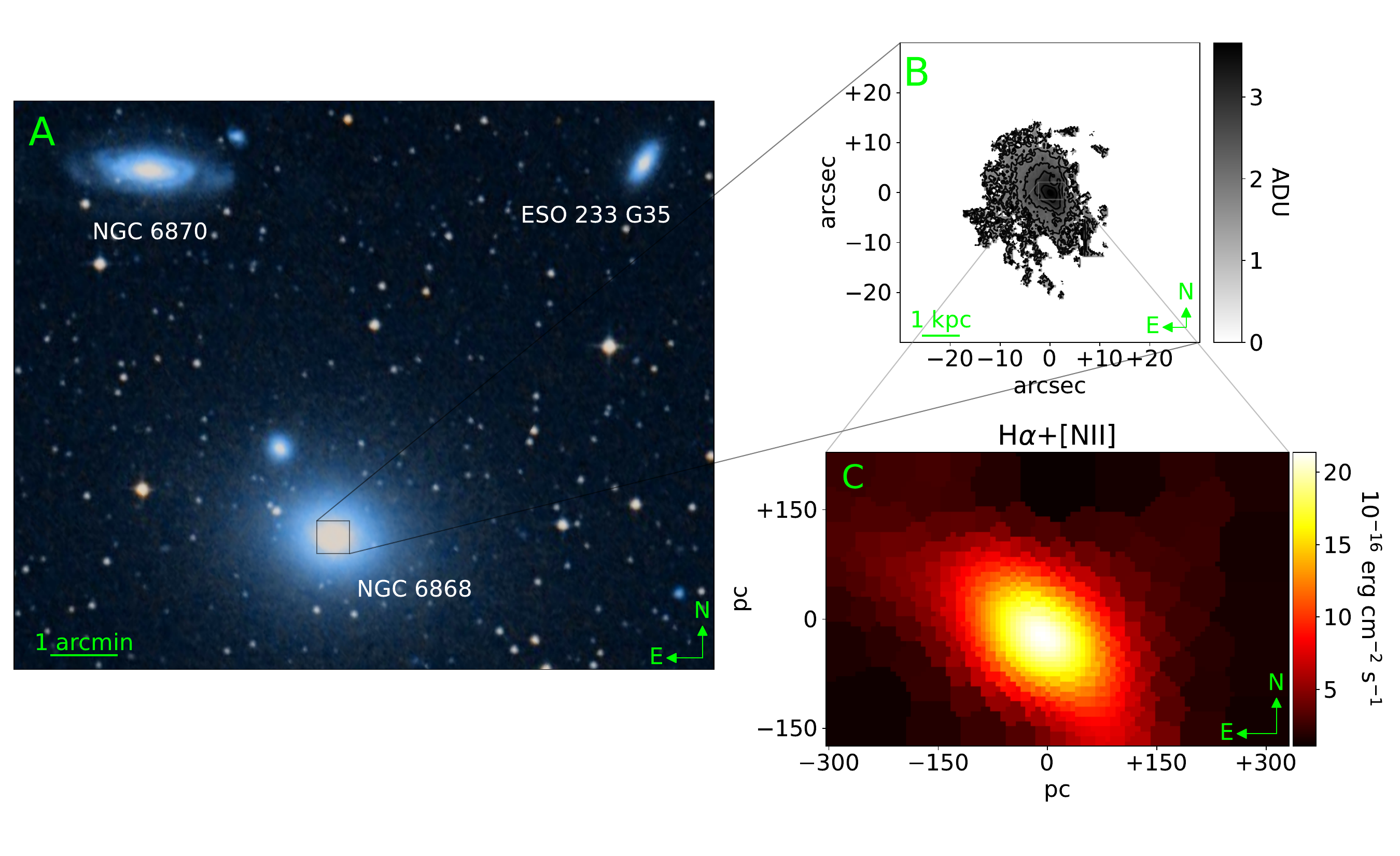}
    \caption{Images from NGC\,6868 in three different scales. (a) Composite DSS image showing NGC\,6868 and close neighbours. It is the brightest group member from the Telescopium group (AS0851). (b) Recreation of H$\alpha$+[\ion{N}{ii}] image present in \citet[][]{MacchettoEtAl1996} using NTT+EFOSC2. They describe the galaxy as having a "regular extended" morphology where small filaments may be present. (c) H$\alpha$+[\ion{N}{ii}] image extracted from the final GMOS data cube after Voronoi binning the data. The (0,0) is kept the same as in Paper~I for consistency.}
    \label{fig:big_picture}
    \end{figure*}

    
    It is clear that to fully understand the nature behind objects classified as LINERs and the impact of an LLAGN in its host galaxy, one needs to engage in detailed spatially resolved studies, tracing both the properties of the ionized gas and the underlying stellar population focused on the region surrounding the SMBH. Therefore, in this series of papers, we present a detailed GMOS IFU study of the object NGC\,6868. It is a nearby \citep[$27.70$~{Mpc},][]{TullyEtAl2013} elliptical galaxy \citep[E2,][]{deVaucouleursEtAl1991}, presents LINER-like signatures and a complex ionization profile \citep[][]{RickesEtAl2008}. We notice that, to the best of our knowledge, no dedicated studies of its central region have been made yet. Table~\ref{tab:basic_pars} shows some basic parameters for NGC6868, while Fig.~\ref{fig:big_picture} shows the galaxy at different scales.
    
    \begin{table}
    \centering
    \caption{Table showing some basic parameters of the galaxy NGC\,6868.}
    \label{tab:basic_pars}
    \begin{tabular}{lc}
    \hline
    Parameter                                                                                               & NGC 6868                                                                                                                                     \\ \hline
    RA (J2000)                                                                                              & 20\textsuperscript{h}09\textsuperscript{m}54\textsuperscript{s}.07                                                                           \\
    Dec. (J200)                                                                                             & -48$^{\circ}$22´46.4´´                                                                                                                        \\
    Morphology\textsuperscript{a}                                                                                              & E3                                                                                                                                           \\
    R (mag)\textsuperscript{b}                                                                                                & 7.91                                                                                                                                         \\
    M$_\text{R}$ (mag)\textsuperscript{b}                                                                                             & -24.7                                                                                                                                        \\
    Diameter (kpc)\textsuperscript{c}                                                                                            & 73.0                                                                                                                                        \\
    L$_\text{X}$ (erg\:s$^{-1}$)\textsuperscript{d}                                                                             & $8.54\cdot10^{40}$                                                                                                                                    \\ 
    Nuclear Activity\textsuperscript{e}                                                                                           & LINER                                                                                                                                    \\ 
    Radio classification\textsuperscript{f}                                                                                    &  Flat-Spectrum Radio Source	                                                                                                                                \\ 
    A$_\text{V}$\textsuperscript{g} (mag)                                                                                      & 0.152                                                                                                                                        \\
    Radial Velocity\textsuperscript{h} (km\,s$^{-1}$)                                                                            & 2854                                                                                                                                         \\
    Distance\textsuperscript{i} (Mpc)                                                                                          & 27.70                                                                                                                                        \\
    Redshift\textsuperscript{h} (z)                                                                                            & 0.00952                                                                                                                                      \\\hline
    \multicolumn{2}{l}{Data available in NED\tablefootnote{The NASA/IPAC Extragalactic Database (NED) is operated by the Jet Propulsion Laboratory, California Institute of Technology, under contract with the National Aeronautics and Space Administration}}\\
    \textsuperscript{a}\citet{deVaucouleursEtAl1991} & \textsuperscript{b}\citet{CarrascoEtAl2006}\hfill\vadjust{}\\
    \textsuperscript{c}\citet{LaubertsValentijn1989} & \textsuperscript{d}\citet{BabykEtAl2018}\hfill\vadjust{}\\
    \textsuperscript{e}\citet{RickesEtAl2008} & \textsuperscript{f}\citet{HealeyEtAl2007}\hfill\vadjust{}\\
    \textsuperscript{g}\citet{SchlaflyFinkbeiner2011} & \textsuperscript{h}\citet{RamellaEtAl1996}\hfill\vadjust{}\\
    \multicolumn{2}{l}{\textsuperscript{i}\citet{TullyEtAl2013}}
    \end{tabular}
    \end{table}
    
    \subsection{About NGC\,6868}

    NGC\,6868 is the brightest member of the Telescopium group and has been already observed in different wavelengths. It was featured in a series of papers investigating the ISM on early-type galaxies through long-slit spectroscopy \citep[e.g.][]{BusonEtAl1993, ZeilingerEtAl1996, MacchettoEtAl1996, PizzellaEtAl1997, FerrariEtAl1999, CaonEtAl2000, FerrariEtAl2002}. They found that the ionized gas in NGC\,6868 has a ``regular extended'' morphology with possible small filaments. Also, the stars present a kinematically decoupled core (KDC) as the stars in the central region counter-rotate with respect to the stars in the outer regions. Moreover, they found that NGC\,6868 harbours complex ionized gas kinematics with the superposition of two ionized gas discs. The hot dust mass estimated from the mid-IR is 70~{M$_\odot$} \citep{FerrariEtAl2002}. \citet[][]{Veron-CettyVeron1988} detected a dust lane in the centre of NGC\,6868 further endorsed by \citet[][]{BregmanEtAl1998} who detected cold dust using IRAS data. \citet[][]{HansenEtAl1991} examined this galaxy using CCD images and an International Ultraviolet Explorer low-resolution spectrum, detecting a dust lane with spiral features emerging from it. They also report that the Ly$\alpha$ distribution follows that of dust, which suggests that NGC6868 has captured a gas-rich galaxy. In fact, \citet{MachacekEtAl2010} using X-ray data found strong evidence of a past encounter between NGC\,6868 and NGC\,6861 in the past hundred Myr, displaying tidal tails and shells. Moreover, they found X-ray cavities, indicative of past AGN activity triggered by the interaction. Radio observations \citep[][]{SleeEtAl1994, MauchEtAl2003, HealeyEtAl2007} revealed a low-power flat spectrum radio source in its centre ($\alpha \sim 0.07$) and the brightness, temperature and spectral slope are inconsistent with HII regions, thus hinting at an AGN as the most likely source of the emission. \citet{RoseEtAl2019} analysed molecular gas in the centre of NGC\,6868 and concluded it is drifting in non-circular motions. More recently, \citet[][]{RicciEtAl2023}, using the same GMOS data presented in this paper, included NGC\,6868 in their analysis of the nuclear region of early-type galaxies, finding compelling evidence for an AGN, despite no detection of a broad-line region. They integrated all the spectra from the NGC\,6868 data cube within the PSF, centred in the stellar photometric peak. In this work, we explored the whole FoV of this observation.
    
    In \citet[][hereafter Paper~I]{BenedettiEtAl2023} we already studied the stellar content of this galaxy through stellar population synthesis and indices measurements. We found that this galaxy is dominated by old stars (\(\sim\)12~{Gyr}) with high-metallicity (1.0-1.6~{Z$\odot$}) with a shallow contribution of a young (\(\sim\)63~{Myr}) also high-metallicity (1.6~{Z$\odot$}). Indices revealed that this object has a complex chemical evolution, further endorsed by the [\textalpha/Fe] map showing regions with very distinct enrichment. The stellar kinematics in the centre of the galaxy is dispersion dominated and no apparent ordered rotation was detected as well as no evidence for a featureless continuum. These findings led us to conclude that this galaxy probably is experiencing only a residual level of star formation. This contribution, however, is not enough to explain the complex enrichment profile which we attribute to past mergers due to the nature of NGC\,6868 (brightest galaxy of the Telescopium group).

    In this paper, we will focus on the ionized gas content of NGC\,6868, organised as follows: in \S~\ref{sec:data}, we describe the observations and the reduction procedures; in \S~\ref{sec:ifscube}, we present the methodology; in \S~\ref{sec:results}, the results are presented. Discussion of the results is made in \S~\ref{sec:discuss} and the conclusions and summary are made in \S~\ref{sec:conclusion}. Throughout this paper, we assume that solar metallicity corresponds to $\text{Z}_\odot=0.019$ \citep[][]{GirardiEtAl2000}.

\section{Observations and data reduction}
\label{sec:data}

    
    The acquisition of the observational data, the reduction processes and subsequent data processing steps were already thoroughly described in Paper~I. Briefly, NGC\,6868 was observed as part of the DIVING\textsuperscript{3D} survey \citep{SteinerEtAl2022} on 2013 May 04 using the Gemini Multi-Object Spectrograph (GMOS) in the IFU mode mounted on the Gemini South Telescope. The configuration was a one-slit setup (resulting in a FOV of $3.5 \times 5.0$~{arcsec\textsuperscript{2}}), with the B600-G5323 grating centred at 5620~{\AA}. This results in a spectral baseline covering 4260 - 6795~{\AA}. The spectral resolution was estimated with the \ion{O}{i} $\lambda5577$ line resulting in 1.8~{\AA} (FWHM). Using the acquisition image from GMOS in the \textit{r}-band and field stars present, the seeing was measured at 0.77~{arcsec}. Lastly, the DA white dwarf EG 274 \citep{HamuyEtAl1992} was observed to perform the spectrophotometric calibrations. Standard IRAF procedures \citep{Tody1986, Tody1993} were employed to reduce the data using the {\sc Gemini iraf} package and the {\sc lacos} software \citep{vanDokkum2001} was used to remove cosmic rays. The final sampling of the data cube was 0.05~{arcsec}.
    
    Other data treatments were applied to improve data visualisation as described in \citet{MenezesEtAl2019}: the removal of high-frequency spatial noise using a Butterworth filter \citep{GonzalezWoods2008, RicciEtAl2014}
    ; the correction by the differential atmospheric refraction
    ; and the PCA Tomography technique \citep[][and references therein]{SteinerEtAl2009} was applied to remove instrumental fingerprints. The data cube was then corrected by the Galactic reddening using the CCM law \citep{CardelliEtAl1989} and assuming $A_V=0.152$~{mag} \citep[][]{SchlaflyFinkbeiner2011}. Telluric lines were removed and the spectra were brought to rest-frame velocities using the redshift from \citet{RamellaEtAl1996} ($z = 0.00952$). To improve the PSF from our data, the Richardson-Lucy deconvolution \citep{Richardson1972, Lucy1974} was applied reaching a PSF of 0.71~{arcsec}. 
    The final spatial coverage of the data cube in the source corresponds to an FoV of (\(\sim\)$680\times470$~{pc\textsuperscript{2}}). In Fig.~\ref{fig:big_picture}, a map from [\ion{N}{ii}]+H$\alpha$ is shown, extracted after the subtraction of the stellar content and the subsequent Voronoi binning of the data.
    

\section{Methods}
    \label{sec:ifscube}

    \subsection{Pre-processing: Voronoi binning and continuum fitting}
    
    As mentioned, we have already analysed the stellar populations of this galaxy in Paper~I. In order to analyse the pure emission line spectrum, we subtracted the modelled stellar continuum from our data. To improve the S/N mainly towards the boarders of the FoV, we used the Voronoi binning technique \citep[][]{CappellariCopin2003} which bins the data preserving the maximum spatial resolution given a minimum S/N to be achieved. We measured the signal as the mean flux density between 6528-6615~{\AA} which encompasses the [\ion{N}{ii}] $\lambda\lambda6548,6584$+H$\alpha$ emission. The noise was estimated as the standard deviation between 4800-4845~{\AA}. We opted to use this wavelength range because it is on the bluer side of our data and therefore is noisier, functioning as an upper limit to the noise in the whole spectrum. For this work, we set the threshold to S/N=30 so that no binned spaxels would be greater than the seeing. In Fig.~\ref{fig:regions}, the H$\alpha$ flux map after binning can be seen. 
    
    The stellar synthesis method, despite its precision, can lead to discrepancies between the data and the stellar continuum. This can be worked around by fitting a high-degree polynomial to model a non-physical continuum that can affect our measurements. We masked the emission lines and fitted the continuum using double the weight in arbitrary continuum bands of 20~{\AA} of width surrounding each line. We took extra care to avoid the regions of absorption or emission lines within these bands. We tested different configurations for the fit, varying the weights in each window or masking certain parts of the spectrum, the degree of the polynomial fitted, and selecting a 13th-degree polynomial as the most adequate for the pseudo-continuum. This pseudo-continuum accounts for template mismatch issues or limitations in the stellar population base used to fit the stellar continuum (e.g. abundance ratio variations). Therefore, at first glance, a 13-degree polynomial seems redundant for a polynomial fit, but considering the number of points in the wavelength range ($>2000$ points) coupled with the number of unknowns previously enumerated, we suggest this is an adequate degree. Moreover, when polynomials with smaller degrees were used, the bluest and reddest wavelengths suffered from spurious features created in the pseudo continuum. In Fig.~{\ref{fig:lines}}, we show an example of a continuum fit.

    \subsection{Emission line fitting}
    After this pre-processing, we modelled the emission line using {\sc ifscube} package \citep[][]{Ruschel-DutraOliveira2020}\footnote{Available at: https://github.com/danielrd6/ifscube}, which fits the different emission lines using Gaussian functions with predefined components. 
    
\begin{figure*}
    \centering
    \includegraphics[width=1.0\textwidth, keepaspectratio]{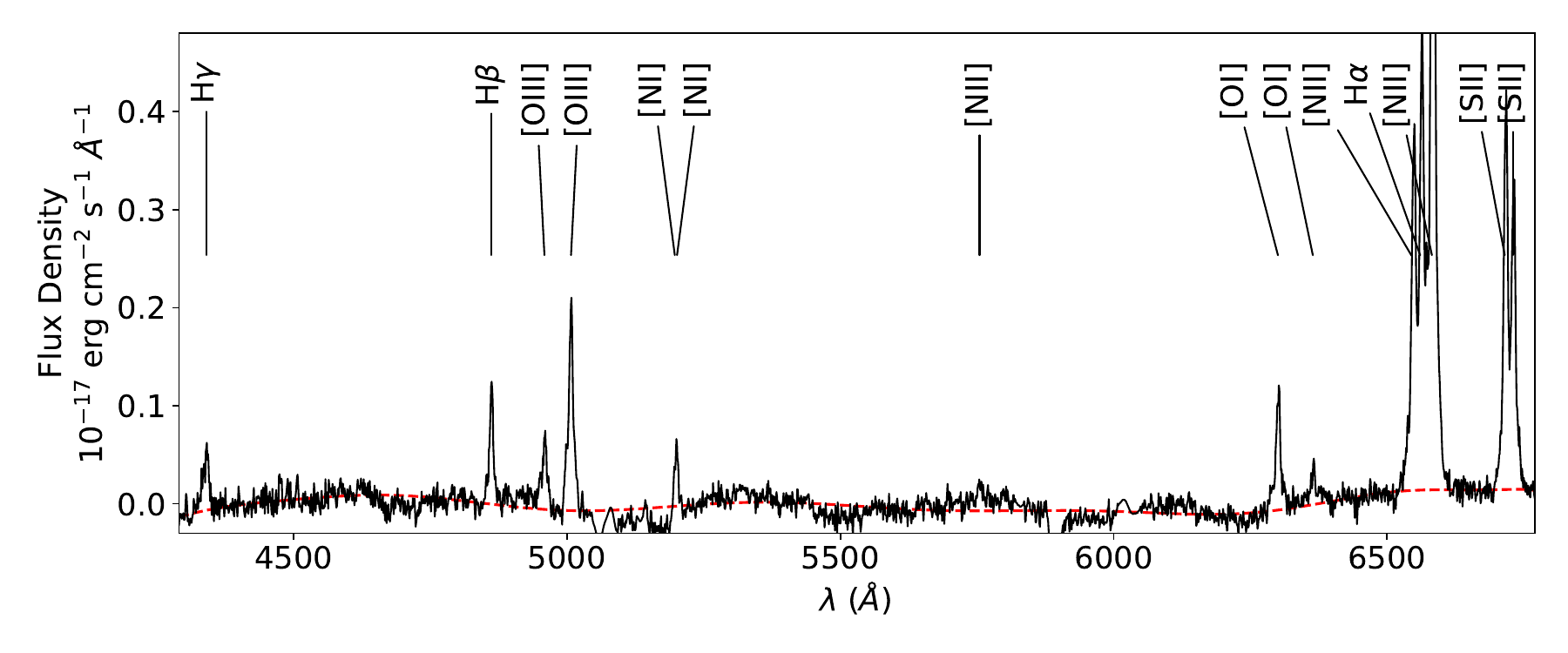}
    \caption{In black, Absorption line free emission spectrum extracted at the peak of the continuum for an aperture of $0.05\times0.05$~{arcsec$^2$}. The fitted pseudo-continuum is displayed in red. The most prominent emission lines are identified. More details of the kinematical nature of the lines can be seen in Fig.~\ref{fig:regions}. }
    \label{fig:lines}
    \end{figure*}

\begin{figure*}
    \centering
    \includegraphics[width=1.0\textwidth, keepaspectratio]{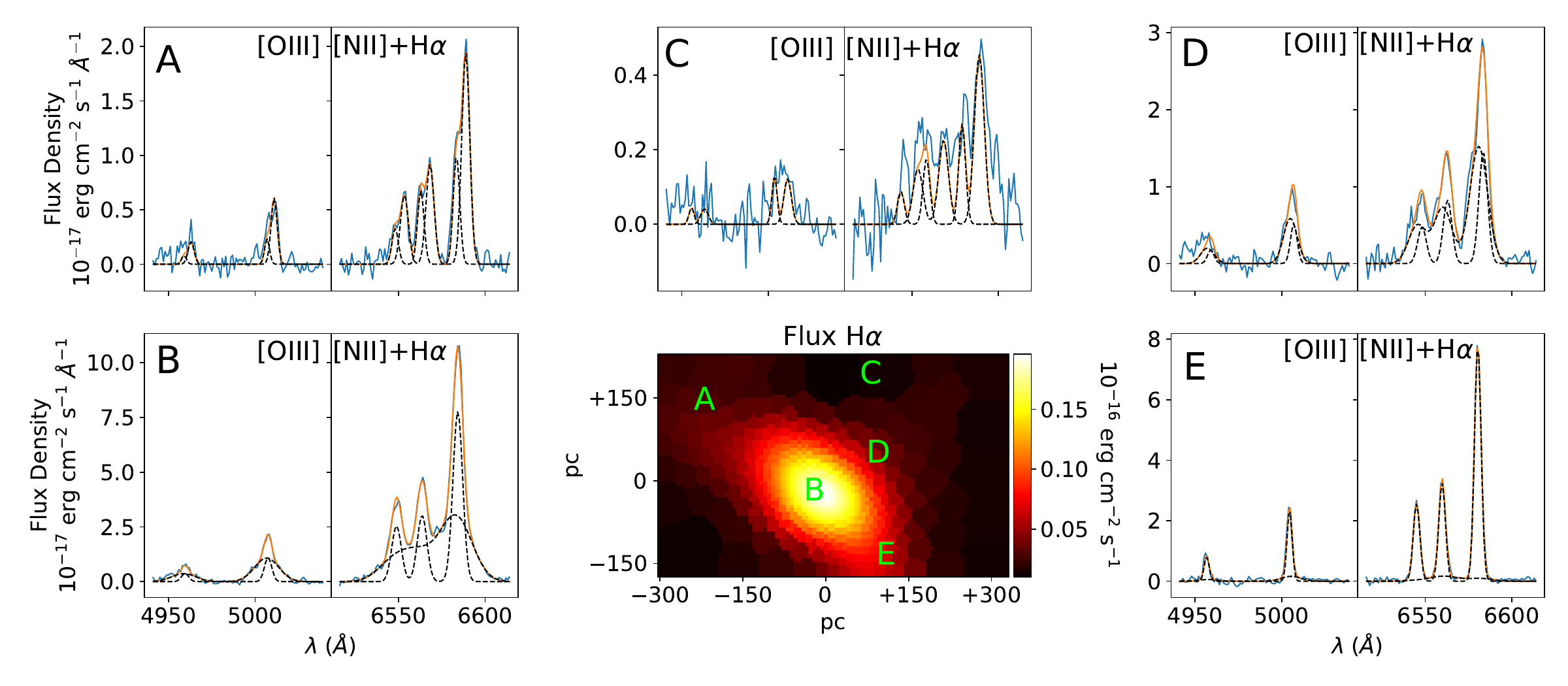}
    \caption{Different regions and the respective line profiles from each assigned Voronoi bin. In blue, the observed profile of the [\ion{O}{iii}] (left) and H$\alpha$+[\ion{N}{ii}] (right), in dashed black lines, each modelled Gaussian and, in orange, the total model. The huge diversity among the line profiles becomes evident, showing double peaks, broad, narrow and wing components which our model successfully reproduces. The central bottom panel shows the total H$\alpha$ flux estimated through the fitting procedure.}
    \label{fig:regions}
    \end{figure*}

    The spectrum from NGC\,6868 is rich in emission lines as can be seen in Fig.~\ref{fig:lines}. The properly fitted lines are \ion{H}{i} H$\alpha$, H$\beta$ and H$\gamma$, [\ion{N}{II}]~$5755, 6548, 6583$~{\AA}, [\ion{O}{iii}]~$4959, 5007$~{\AA}, [\ion{O}{i}]~$6300, 6360$~{\AA}, [\ion{N}{i}]~$5197, 5200$~{\AA} and [\ion{S}{ii}]~$6716, 6731$~{\AA}. By looking at Fig.~\ref{fig:regions}, the immense diversity in line profiles from this galaxy becomes clear. We carried out a series of tests with different configurations, such as varying the number of components in each line, adding a broad component compatible with what we would expect for a broad-line region in the \ion{H}{I} recombination lines, fitting each section of the spectra separately, removing or adding kinematical constraints (e.g. coupling or not the [\ion{O}{iii}] kinematics with the other lines). Looking at the residuals and using the model with the least number of components, we ended up with a model consisting of two distinct kinematical components for each line: a narrow ($\sigma \sim 80-130$~{km\:s$^{-1}$}) and a broad ($\sigma \sim 100-450$~{km\:s$^{-1}$}) component. This has been applied for all the spaxels and some example fits can be seen in Fig.~\ref{fig:regions}.
    
    In order to reduce the degeneracy from our fit, we made some assumptions and established some constraints to our models. Firstly, we used the following well-known line ratios \citep[][]{OsterbrockFerland2006}: f\textsubscript{[\ion{N}{ii}]~$\lambda6583$}~= 3.06$\cdot$f\textsubscript{[\ion{N}{ii}]~$\lambda6548$}, f\textsubscript{[\ion{O}{iii}]~$\lambda5007$}~= 2.94$\cdot$f\textsubscript{[\ion{O}{iii}]~$\lambda4959$}, f\textsubscript{[\ion{O}{i}]~$\lambda6300$}~= 3.05$\cdot$f\textsubscript{[\ion{O}{i}]~$\lambda6360$}. On the first try, we only established kinematical groupings with components from lines produced by the same ion. We noticed, however, that all lines ended up with really similar kinematical and flux distributions among different ions. Despite typically behaving differently when compared to the other lines, this is valid even for the [\ion{O}{iii}] lines. Secondly, we constrained that each component from all the lines fitted would be in the same kinematical group e.g. the blueshifted component needs to have the same velocity and velocity dispersion among all the different lines. This helps to disentangle the different components in weaker lines as the kinematical information from stronger lines is used to improve the fit. 
    
    Electron temperature ($T_e$) and density ($n_e$) were computed using {\sc PyNeb} \citep[][]{LuridianaEtAl2015} with line-ratios [\ion{N}{ii}]~$\lambda5755/\lambda6583$ and [\ion{S}{ii}]~$\lambda6716/\lambda6731$, respectively. We also calculated the emission line ratios of [\ion{N}{ii}]~$\lambda6583$/H$\alpha$, [\ion{S}{ii}]~$\lambda6716, 6731$/H$\alpha$, [\ion{O}{i}]~$\lambda6300$/H$\alpha$ , [\ion{O}{iii}]~$\lambda5007$/H$\beta$. We derived the reddening also using the {\sc PyNeb} package using the ratio between the H$\alpha$ and H$\beta$. Considering the case B recombination and rough estimates of both density and temperature, as a first approximation (100~{cm\textsuperscript{-3}}; 10000~{K}). Using this parameters, we estimate the theoretical value of $\text{F}_{\text{H}\alpha}/\text{F}_{\text{H}\beta} = 2.87$. Assuming, a CCM extinction law \citep[$f_\lambda$;][R\textsubscript{V}=3.1]{CardelliEtAl1989} and following \citet[][]{RiffelEtAl2021}, we get
    \begin{align}
    \text{E}(\text{B}-\text{V}) &= \frac{E(\text{H}\beta-\text{H}\alpha)}{f_\lambda(\text{H}\beta)-f_\lambda(\text{H}\alpha)}\\
                  &= \frac{2.5}{3.1\cdot(1.164-0.818)}\log\left(\frac{(\text{F}_{\text{H}\alpha}/\text{F}_{\text{H}\beta})^{\text{obs}}}{(\text{F}_{\text{H}\alpha}/\text{F}_{\text{H}\beta})^{\text{theo}} }\right) \notag \\
    \text{A}_\text{V} &=7.22\log\left(\frac{(\text{F}_{\text{H}\alpha}/\text{F}_{\text{H}\beta})^{\text{obs}}}{2.87}\right).
    \label{eq:vel_proj}
    \end{align}
    
    Therefore, using the Balmer recombination lines (H$\alpha$, H$\beta$), we derived the reddening in the V band (A\textsubscript{V}) and deredden all emission lines, following
    \begin{equation}
        \text{F}_\text{int}=\text{F}_\text{obs}10^{0.4\cdot\text{A}_\lambda}=\text{F}_\text{obs}10^{0.4\cdot\text{A}_\text{V}\cdot f_\lambda}.
    \end{equation}

\section{Results}
\label{sec:results}
 
 \subsection{Gas kinematics and distribution}
    
    \subsubsection{Channel Maps}
    
\begin{figure*}
    \centering
    \includegraphics[width=1.0\textwidth, keepaspectratio]{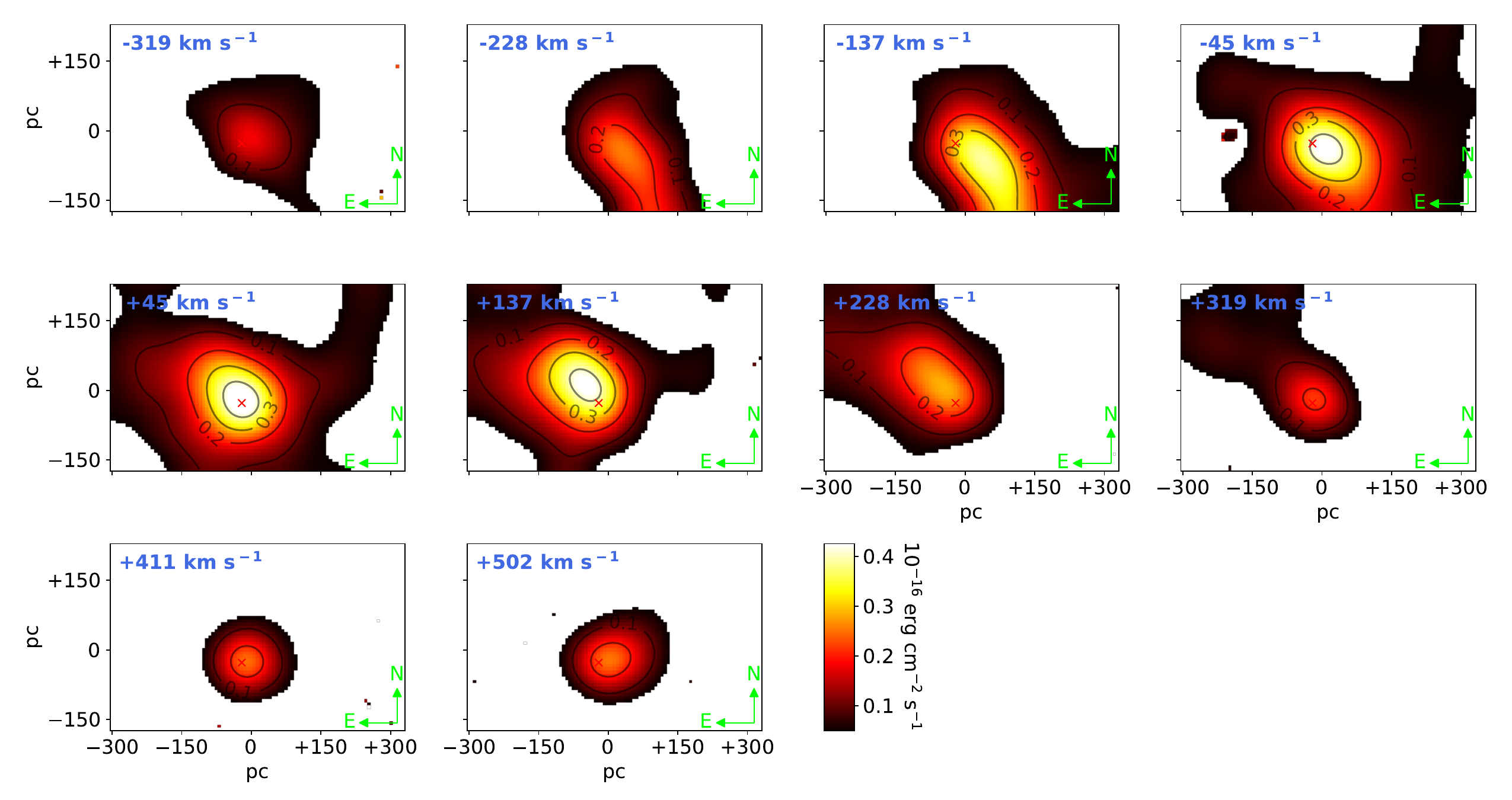}
    \caption{Channel maps for the H$\alpha$ emission line using 2.0~{\AA} (\(\sim\)91~{km\:s$^{-1}$}) as our sampling width. We masked values smaller than three times the standard deviation as calculated in \S~\ref{sec:ifscube}. In the last map we can already see a contamination from the surrounding [\ion{N}{ii}]~$6583$~{\AA}.}
    \label{fig:channel_maps}
    \end{figure*}

    We slice the data cube in steps of 2.0~{\AA} (\(\sim\)91~{km\:s$^{-1}$}) in order to create several H$\alpha$ lines maps (Fig.~\ref{fig:channel_maps}). In this way, we can correlate a spatial counterpart from the gas with a particular kinematical signature. We masked all spaxels that had flux values smaller than 3 times the value of the standard deviation as calculated in \S~\ref{sec:ifscube} for that spaxel. We have applied the same procedure for the [\ion{O}{iii}]~$5007$~{\AA} line and have not found significant differences between both lines. 
    
    The central region of NGC\,6868 displays several distinct features. In the high-velocity maps ($v=+411$ and $+502$~{km\:s$^{-1}$}), a circular distribution is seen, with its centre coinciding with that estimated on Paper~{I}. In the lowest velocity map ($v=-319$~{km\:s$^{-1}$}), the same centrally concentrated region appears albeit with a slight perturbation to the NW that can be followed in the maps of $v=-228$ and $-137$~{km\:s$^{-1}$}. In the maps from $v=-228$ to $+228$~{km\:s$^{-1}$}, a signature of a disc can be traced, going from the SW to NE. In the $v=+319$~{km\:s$^{-1}$} a distortion in the central profile can also be seen in the NE region, which appears to be independent of the disc as such a profile does not emerge in the $v=-319$~{km\:s$^{-1}$} map. Interestingly, a region in the N section does not seem to have any detected ionized gas, which is better seen in the $v=+45$~{km\:s$^{-1}$} map. A closer inspection of the upper panel of Fig.~\ref{fig:big_picture} reveals that this feature is also seen in the observation of \citet{MacchettoEtAl1996}, showing a slight decrease in flux right in the N border of our FoV when compared to the surrounding region.

    \subsubsection{Velocity and velocity dispersion maps}
   
\begin{figure*}
    \centering
    \includegraphics[width=0.95\textwidth, keepaspectratio]{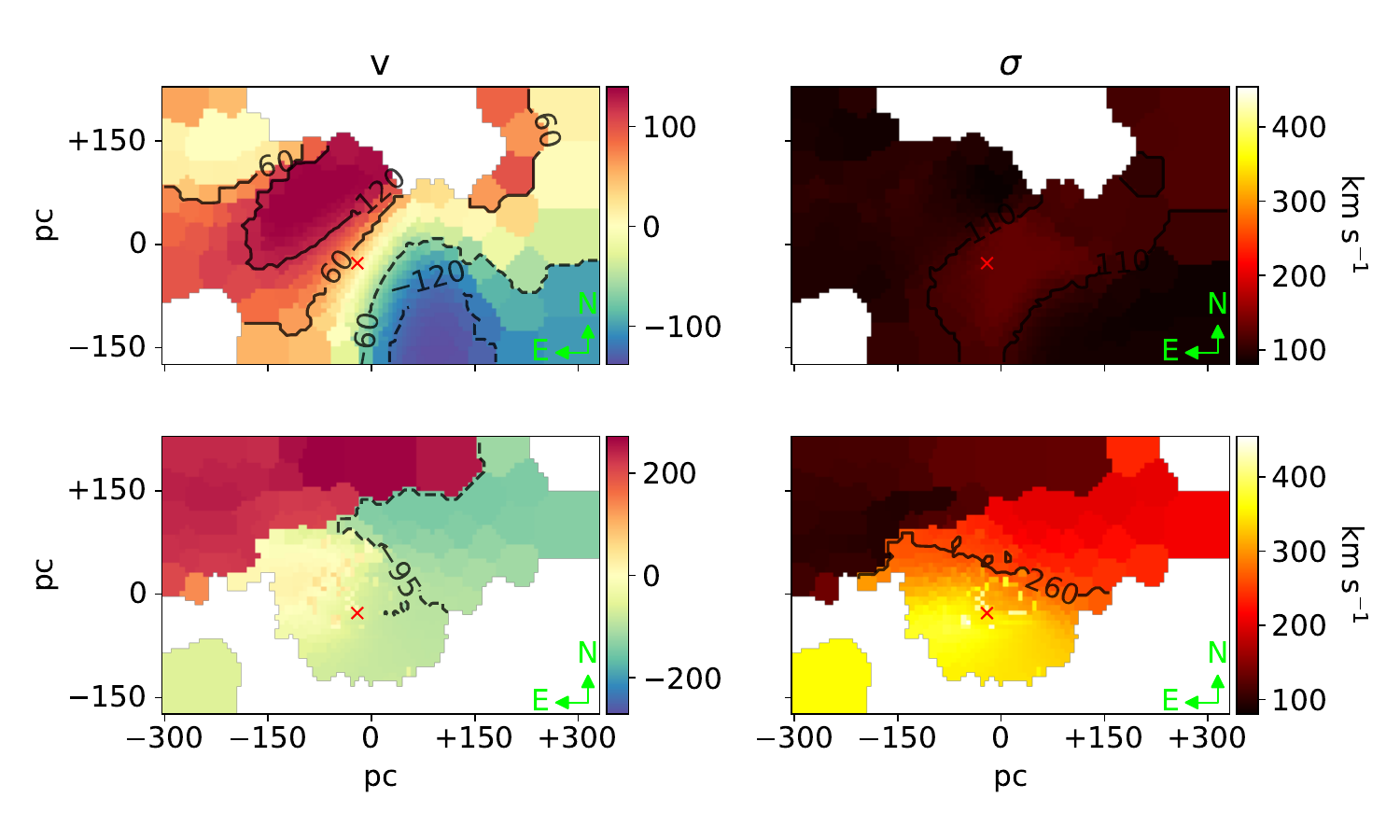}
    \caption{Kinematical results for the central region of NGC\,6868. The top panels show the results for the narrow component and the bottom panels, for the broad component. For consistency, we will maintain this scheme in all figures unless specified. Centroid velocities are displayed on the left and velocity dispersion, on the right.}
    \label{fig:kinematics}
    \end{figure*}

    The centroid velocities and velocity dispersion for the narrow and broad components are presented in Fig.~\ref{fig:kinematics}. It is clear from this figure that the narrow component ranges from \(\sim\)$140$~{km\:s$^{-1}$} to \(\sim\)$90$~{km\:s$^{-1}$}, broadening in the central regions. Looking at the centroid velocities of this component, one can infer that the central area of NGC\,6868 hosts an ionized gas disc, with velocity amplitude ranging from $\pm150$~{km\:s$^{-1}$}. The centre of the rotation profile appears to coincide with the centre of NGC\,6868, estimated in Paper~I. The broader component appears to have a more diverse nature, with regions with a similar velocity dispersion as the narrow component (NE region) but higher centroid velocities (\(\sim\)$250$~{km\:s$^{-1}$}). In the central part, the velocity dispersion reaches\(\sim\)$400$~{km\:s$^{-1}$}. We also notice that in the NW direction, a blueshifted (approximately $-100$~{km\:s$^{-1}$}) broader \(\sim\)$250$~{km\:s$^{-1}$} wing is apparent.
    
    \citet[][]{CaonEtAl2000} have reported a rotating gas disc in NGC\,6868. Our detection agrees with the orientation reported in their study. However, they find profiles peaking at \(\sim\)180 and 190~{km\:s$^{-1}$}, slightly larger than reported here. As their modelling relied on only one Gaussian, they are probably having contamination of the high-velocity components that we can detect separately, mainly in PA=120$^{\circ}$, where they report a counter-rotating gas disc which likely comes from the blueshifted broad component in the NW affecting their measurements. 

    \subsubsection{Flux distributions}
    
\begin{figure*}
    \centering
    \includegraphics[width=0.95\textwidth, keepaspectratio]{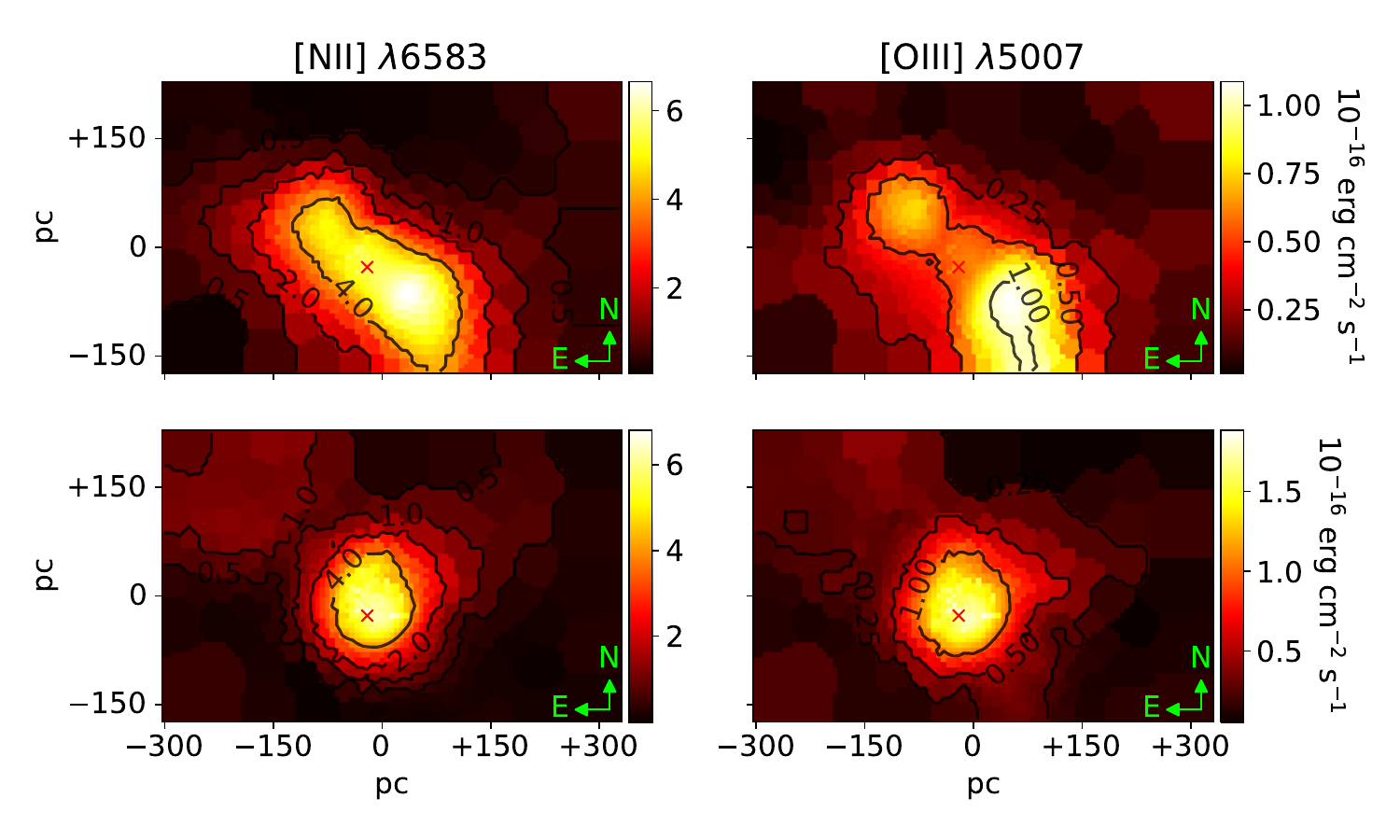}
    \caption{Flux distribution from narrow (top) and broad components (bottom) and [\ion{N}{ii}]~$6583$~{\AA} (left) and [\ion{O}{iii}]~$5007$~{\AA} (right) emission lines not corrected by dust extinction. The components are fairly similar between the different lines. }
    \label{fig:fluxes}
    \end{figure*}

    The flux distributions derived are presented in Fig.~\ref{fig:fluxes}. We show only the [\ion{N}{ii}]~$6583$~{\AA} and [\ion{O}{iii}]~$5007$~{\AA} emission lines flux distributions because [\ion{N}{ii}] is the most prominent line in our cube. Moreover, most lines follow its distribution except for [\ion{O}{iii}]~$5007$~{\AA} which presents its peculiarities, as shown.

    The narrow component distribution appears in a flat distribution along the SW-NE direction, following the rotation profile already described. In the [\ion{O}{iii}] line, however, it appears that the SW region is enhanced. This does not appear to be a feature from the fitting procedure, but rather some physical process enhancing it at this location. This can be roughly seen in Fig.~\ref{fig:regions} as it appears that, in region E, [\ion{O}{iii}] has the largest intensity within all the spectra shown. The broader component shows a distribution that departs from the narrow one with a clear centrally concentrated flux distribution with two wings: one at NE and another at NW, detected in both [\ion{O}{iii}] and [\ion{N}{ii}] lines. This coincides with the regions described also detected in the channel maps and the kinematical maps.

    \subsection{Ionized gas physical parameters}
    
\begin{figure}
    \centering
    \includegraphics[width=1.0\columnwidth, keepaspectratio]{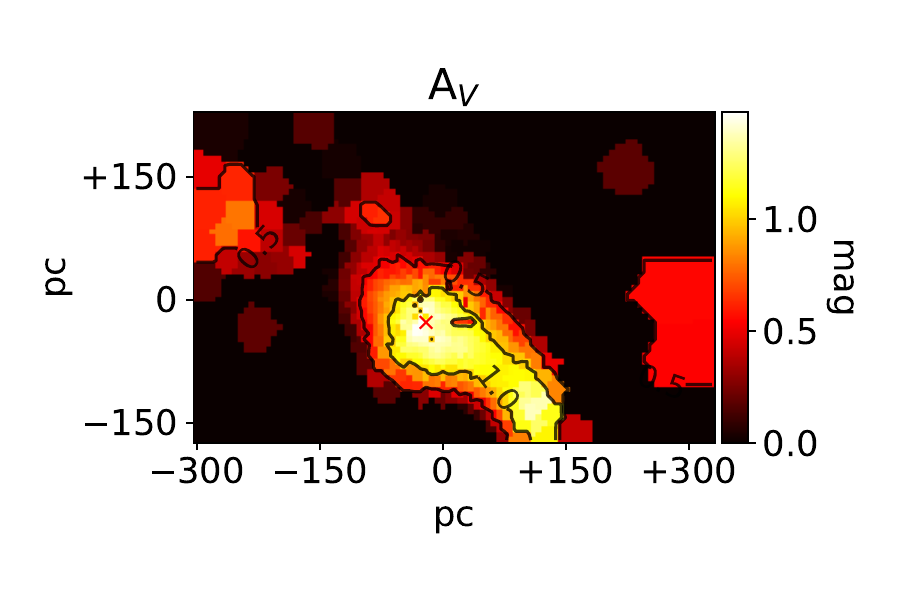}
    \caption{The reddening in the V band derived through the Balmer recombination lines ratio H$\alpha$/H$\beta$. A dust lane emerges in the centre, agreeing with past results found through stellar population synthesis.}
    \label{fig:av_gas}
    \end{figure}

    Using the Balmer recombination lines H$\alpha$ and H$\beta$, we derived the reddening in the V-band (A\textsubscript{V}). We decided to sum the fluxes from both components of each line to measure the A\textsubscript{V} as H$\beta$ is a rather weak emission line in this object. Thus, having a single measurement of this property in each spaxel. We assumed Case B recombination and, as a first approximation, set the temperature and density respectively as 10000~{K} and 100~{cm\textsuperscript{3}}. Using the CCM extinction law \citep[][]{CardelliEtAl1989}, the obtained A\textsubscript{V} map is shown in Fig.~\ref{fig:av_gas}.
    
\begin{figure}
    \centering
    \includegraphics[width=1.0\columnwidth, keepaspectratio]{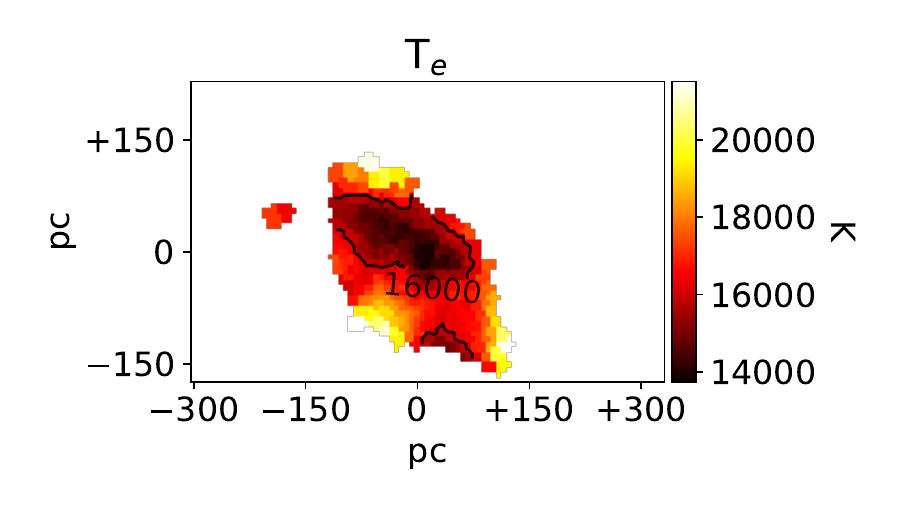}
    \caption{Temperature profile derived from the [\ion{N}{ii}]~$6583,5755$~{\AA} line ratio corrected by the extinction. The temperature profile seems to display a negative gradient with values ranging from 14000 to 21000~{K}.}
    \label{fig:temp}
    \end{figure}

\begin{figure}
    \centering
    \includegraphics[width=1.0\columnwidth, keepaspectratio]{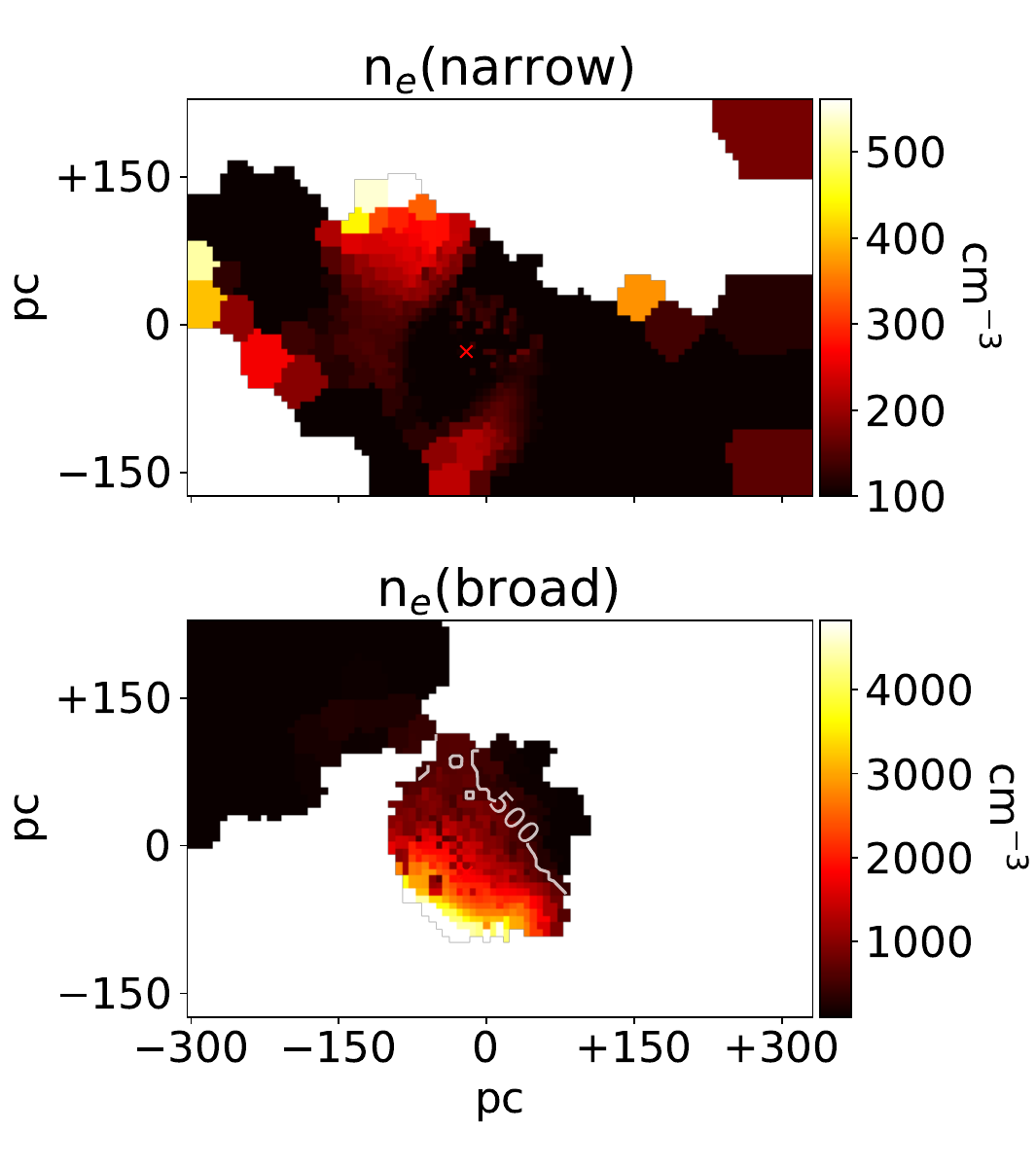}
    \caption{Electron density profile derived from [\ion{S}{ii}]~$\lambda 6716/\lambda6731$ corrected by extinction. \textbf{Top panel:} Density values for the narrow component with values ranging from 100~{cm\textsuperscript{-3}} to 550~{cm\textsuperscript{-3}}. \textbf{Bottom panel:} Broader component showing a large variation in the derived density values. In the central part of the FoV electron density is over 4000~{cm\textsuperscript{-3}} reaching similar values to the narrow component in the NE and NW region. A 500~{cm\textsuperscript{-3}} contour line has been added to aid the visualisation}
    \label{fig:dens}
    \end{figure}

    The A\textsubscript{V} map reveals a dust lane that has already been reported in previous studies \citep[e.g.][]{Veron-CettyVeron1988, HansenEtAl1991, BregmanEtAl1998, FerrariEtAl1999} and also in Paper~I, having the same orientation (PA\(\sim\)$120^{\circ} $) and roughly the same spatial extension. The reddening, however, is enhanced when compared to these previous studies. In Paper~I, we found that the stellar dust distribution peaks at \(\sim\)$0.65$~{mag} whereas the gas A\textsubscript{V} derived here reaches \(\sim\)$1.5$~{mag}. This has been found in past studies, such as \citet[][]{RiffelEtAl2021}, with a ratio close to the one reported here. In that work, it was suggested that this was due to single extinction laws used in {\sc starlight}, the fitting code used \citep{CidFernandesEtAl2005}. Newer generations of stars, however, are typically embedded in the dust reminiscent of the star-formation processes. Thus, the stellar synthesis can underestimate the dust content mainly if old stellar populations dominate the galaxy. Using the same reasoning, the ionising radiation might come from a heavily dust-obscured region, not necessarily young stars, such as an AGN, or the dustier gas ejected during the stellar evolution process, explaining the different line ratios observed. Using the extinction-corrected emission line fluxes, we derived both the electron temperature and density, respectively shown in Fig.~\ref{fig:temp} and Fig.~\ref{fig:dens} which the code computes in an interactive fashion using the result from one parameter to estimate the other until they converge.
    
    We were not able to find any other measurement in the literature of the electron temperature and density for NGC\,6868, making this the first for this object in the literature. Indeed, spatially resolved measurements of the electron temperature in AGN hosts are available only for a few objects, as they require the detection of typically weak emission lines in AGN spectra \citep[][]{RevalskiEtAl2018, RiffelEtAl2021a, RiffelEtAl2021b, NegusEtAl2023}. In Fig.~{\ref{fig:temp}}, the temperature profile of the ionized gas is co-spatial with the broader component flux distribution. Also, it appears that an outward gradient emerges, ranging from 14000~{K} to nearly 21000~{K} at the edges of our detection.
    
    We also provide density estimates for the whole FoV separating the narrow and broad components in the [SII]~$6716,6731$~{\AA}. In some regions, the [SII]~$6716,6731$~{\AA} ratio is larger than 1.45 and density estimates are no longer valid. As this implies a low-density environment, when values exceeded 1.45, we set 100~{cm\textsuperscript{-3}} as the density in that spaxel.
    
    For the narrow component, a ubiquitous low-density component is found, with values ranging from the lower limit 100~{cm\textsuperscript{-3}} to less than 500~{cm\textsuperscript{-3}}. This is not the case for the broader component with values similar to the ones found in the narrow component for the NE region and values in the central region reaching over 4000~{cm\textsuperscript{-3}}. \citet[][]{HansenEtAl1991} based on data from \citet[][]{BonattoEtAl1989} has estimated the electron density to be 800~{cm\textsuperscript{-3}}. As a kinematical decomposition was not carried out in those studies, the in-between value most likely stems from their treatment of the two components found in this work as one.
    
    We estimate the total ionized gas mass using the relation
    \begin{equation}
        \text{M}_{\text{gas}} \approx 2.4\times10^5\frac{\text{L}_{41}(\text{H}\alpha)}{\text{n}_3} \text{M}_\odot
        \label{eq:ion_mass}
    \end{equation}
    as in \citet[][]{doNascimentoEtAl2019}, where $\text{L}_{41}(\text{H}\alpha)$ is the luminosity of H$\alpha$ in units of $10^{41}$~{erg\:s\textsuperscript{-1}} and $\text{n}_3$ is the electron density in units of $10^3$~{cm\textsuperscript{-3}}. Using the extinction corrected H$\alpha$ flux, the distance from Table~\ref{tab:basic_pars} and previously derived electron density, we calculated the ionized gas mass spaxel by spaxel and then summed over the FoV.
    
    The total mass in our FoV is \(\sim\)$ 9.1 \pm 1.2\times10^5$~M$_\odot$. This, however, is a lower limit as the ionized gas component stretches beyond our FoV, as can be seen in Fig.~\ref{fig:big_picture}. \citet[][]{HansenEtAl1991}, using narrow band data to measure the flux of H$\alpha$, have evaluated the \ion{H}{ii} mass as \(\sim\)$ 2\times10^4$~M$_\odot$, bellow what our estimate suggests. \citet{MacchettoEtAl1996} estimated the ionized gas mass as \(\sim\)$ 5\times10^4$~M$_\odot$ also using H$\alpha$ narrow band data. We attribute this difference to several facts. First, each study uses a different value for the distance to NGC\,6868: 27.70~{Mpc} (this work), 36.8~{Mpc} \citep{HansenEtAl1991} and 48.6~{Mpc} \citep{MacchettoEtAl1996}. Also, both studies have used narrow band images to estimate H$\alpha$ using previously determined [\ion{N}{ii}]~$\lambda6583$/H$\alpha$ ratios, instead of having the fine details that spectroscopy can provide. With a data cube, we were able to separate the [\ion{N}{ii}] and H$\alpha$ emission properly, while the narrow band data may suffer some contamination from the [\ion{N}{ii}] lines. Also, they use a fixed electron density value, which as can be seen in Fig.~\ref{fig:dens}, is a simplification. 
    The reddening correction we apply could also explain the different values as they do not apply such a correction. Also, in \citet{HansenEtAl1991} they report problems with their spectrophotometric standards, leading to a 30\% error in their measurements. Of course, our data is also subject to its problems, mainly regarding the flux calibration as standard stars are sometimes observed on different nights, resulting in $\sim 30$~{\%} error in the absolute flux. Therefore, all these differences might come into play to explain the different results.
    
    \subsection{Emission line diagnostic diagrams}
    
\begin{figure}
    \centering
    \includegraphics[width=1.0\columnwidth, keepaspectratio]{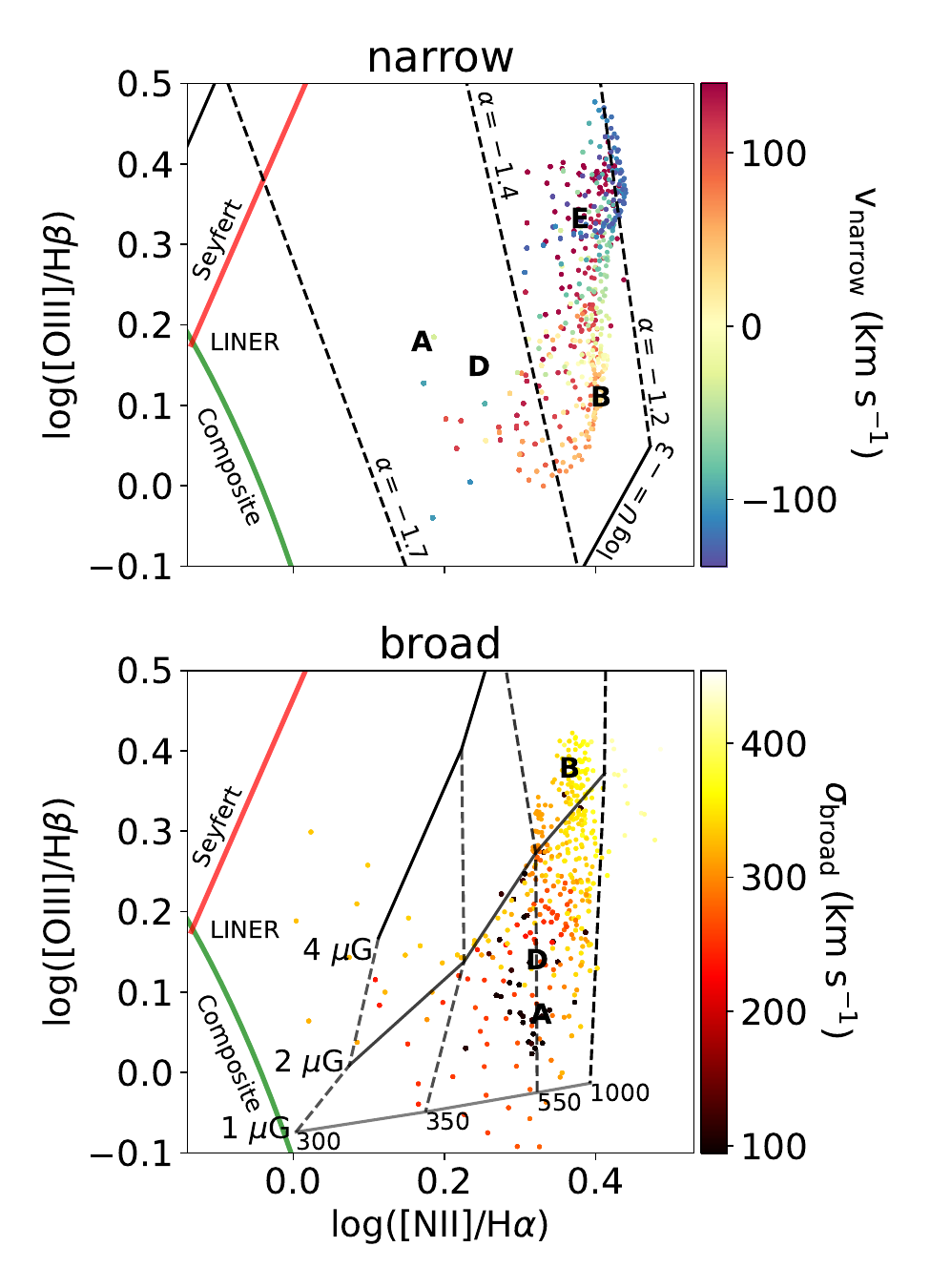}
    \caption{[\ion{N}{ii}]~$\lambda6583$/H$\alpha$ BPT diagram for the central region of NGC\,6868 for the narrow (top panel) and broad (bottom panel). The points were colour-coded following, respectively, the velocity and the velocity dispersion in each component. The whole FoV of our observation is kept within the LINER region, as expected. AGN models from \citet{GrovesEtAl2004} were overplotted in the top panel with $\log U = - 3$ and $\log U = - 2$ (continuous lines) and power-law indices $\alpha=-1.2$, $\alpha=-1.4$ and $\alpha=-1.7$ (dashed lines), as indicated in the plots. Shock grids from \citet[][]{AllenEtAl2008} were overplotted in the bottom panel with models of velocities between 300 and 1000~{km\:s\textsuperscript{-1}} (dashed lines), and magnetic fields between 1.0 and 4.0 $\mu$G (continuous lines). In both panels, we also plot the results obtained in the bins defined in Fig.~{\ref{fig:regions}}, following the same tags.}
    \label{fig:bpt}
    \end{figure}

\begin{figure}
    \centering
    \includegraphics[width=1.0\columnwidth, keepaspectratio]{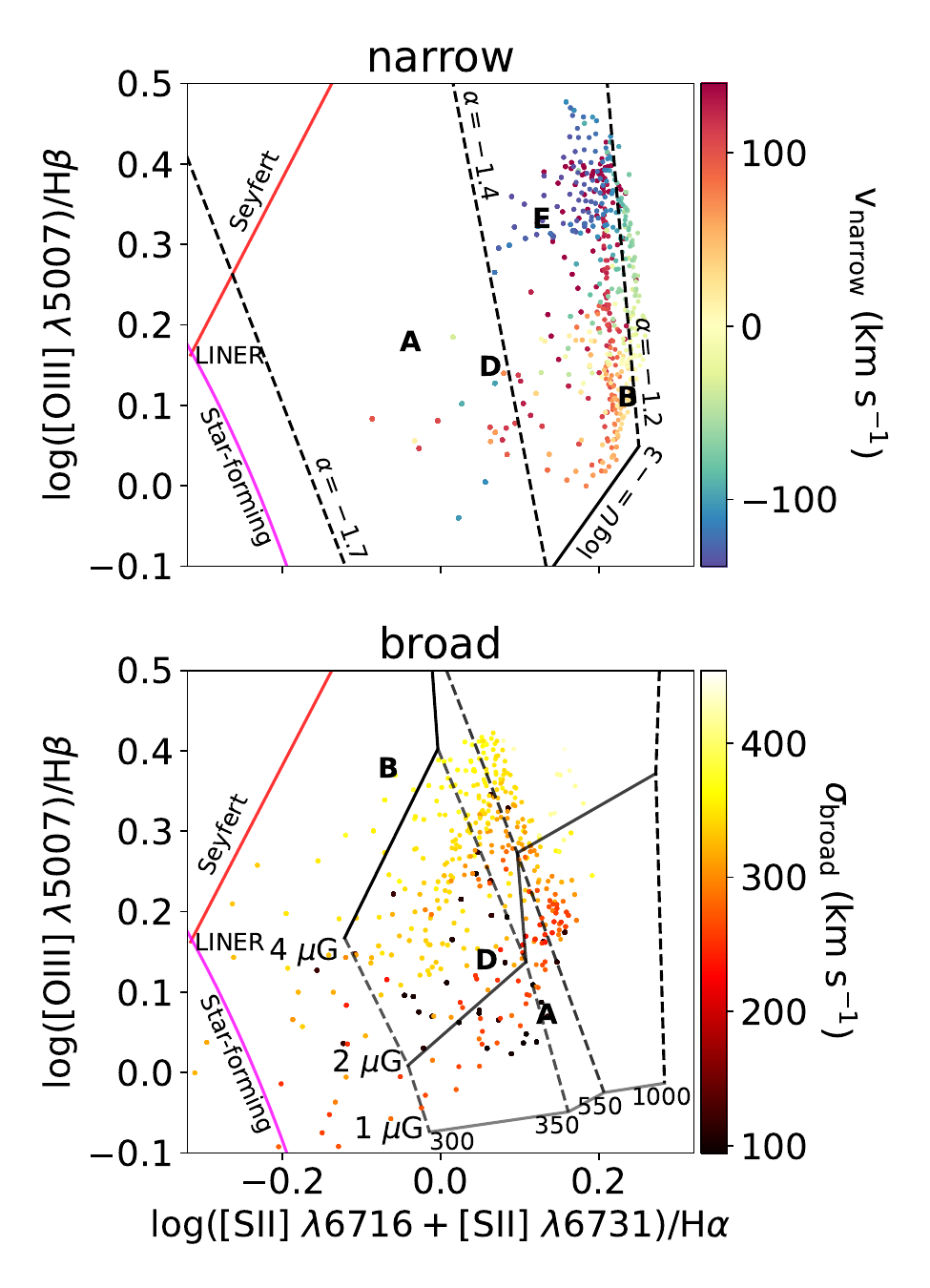}
    \caption{[\ion{S}{ii}]~$\lambda\lambda6716,6731$/H$\alpha$ BPT diagram for the central region of NGC\,6868 for the narrow (top panel) and broad (bottom panel). The points were colour-coded following, respectively, the velocity and the velocity dispersion in each component. The immense majority of the FoV of our observation is kept within the LINER region, as expected. AGN models from \citet{GrovesEtAl2004} were overplotted in the top panel with $\log U = - 3$ and $\log U = - 2$ (continuous lines) and power-law indices $\alpha=-1.2$, $\alpha=-1.4$ and $\alpha=-1.7$ (dashed lines), as indicated in the plots. Shock grids from \citet[][]{AllenEtAl2008} were overplotted in the bottom panel with models of velocities between 300 and 550~{km\:s\textsuperscript{-1}} (dashed lines), and magnetic fields between 1.0 and 4.0 $\mu$G (continuous lines). In both panels, we also plot the results obtained in the bins defined in Fig.~{\ref{fig:regions}}, following the same tags.}
    \label{fig:bpt_sii}
    \end{figure}

\begin{figure}
    \centering
    \includegraphics[width=1.0\columnwidth, keepaspectratio]{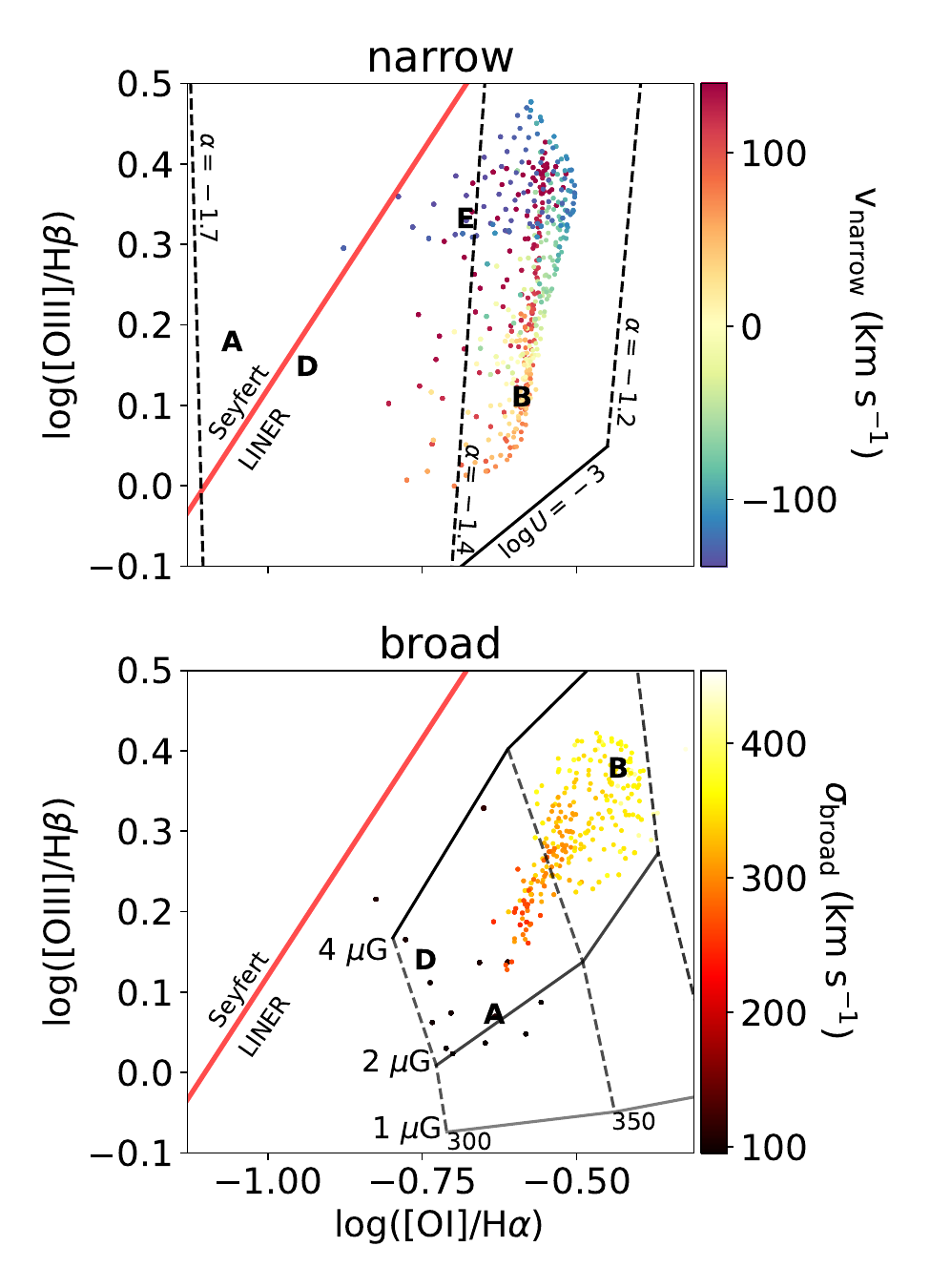}
    \caption{[\ion{O}{i}]~$\lambda6300$/H$\alpha$ BPT diagram for the central region of NGC\,6868 for the narrow (top panel) and broad (bottom panel). The points were colour-coded following, respectively, the velocity and the velocity dispersion in each component. The immense majority of the FoV of our observation is kept within the LINER region, as expected. AGN models from \citet{GrovesEtAl2004} were overplotted in the top panel with $\log U = - 3$ and $\log U = - 2$ (continuous lines) and power-law indices $\alpha=-1.2$, $\alpha=-1.4$ and $\alpha=-1.7$ (dashed lines), as indicated in the plots. Shock grids from \citet[][]{AllenEtAl2008} were overplotted in the bottom panel with models of velocities between 300 and 550~{km\:s\textsuperscript{-1}} (dashed lines), and magnetic fields between 1.0 and 4.0 $\mu$G (continuous lines). In both panels, we also plot the results obtained in the bins defined in Fig.~{\ref{fig:regions}}, following the same tags.}
    \label{fig:bpt_oi}
    \end{figure}

    In order to assess the ionization mechanism present in the central region of NGC\,6868, diagrams discerning the different ionization sources are needed. The most widely used diagrams in the literature are the BPT diagrams and rely on the [\ion{N}{ii}]~$\lambda6583$/H$\alpha$ vs. [\ion{O}{iii}]~$\lambda5007$/H$\beta$ \citep[][]{BaldwinEtAl1981}, [\ion{S}{ii}]~$\lambda6716,6731$/H$\alpha$ vs. [\ion{O}{iii}]~$\lambda5007$/H$\beta$ \citep{VeilleuxOsterbrock1987} and [\ion{O}{i}]~$\lambda6300$/H$\alpha$ vs. [\ion{O}{iii}]~$\lambda5007$/H$\beta$ \citep{VeilleuxOsterbrock1987} line ratios to disentangle the possible ionization mechanisms. \citet[][]{KauffmannEtAl2003} and \citet[][]{KewleyEtAl2006} have established calibrations to separate AGNs from LINERs and star-forming galaxies. We thus employed these lines and the corrected fluxes previously derived to create the diagnostic diagram. The BPT diagrams for [\ion{N}{ii}]~$\lambda6583$/H$\alpha$ vs. [\ion{O}{iii}]~$\lambda5007$/H$\beta$, [\ion{O}{i}]~$\lambda6583$/H$\alpha$ vs. [\ion{O}{iii}]~$\lambda5007$/H$\beta$ and [\ion{S}{ii}]~$\lambda6716,6731$/H$\alpha$ vs. [\ion{O}{iii}]~$\lambda5007$/H$\beta$ are seen respectively in Fig.~\ref{fig:bpt}, Fig.~\ref{fig:bpt_sii} and Fig.~\ref{fig:bpt_oi}.
    
    The three diagrams classify the whole FoV as having LINER-like emission. This was expected as other studies already mentioned had detected the same signatures \citep[e.g.][]{RickesEtAl2008}. As mentioned, shock-heated gas may be present in NGC\,6868. Therefore shock models from \citet[][]{AllenEtAl2008} were overplotted in the broad lines BPT. We used twice-solar metallicity models and a pre-shock density of 1~{cm\textsuperscript{-3}}, creating our grid with models of velocities between 300 and 1000~{km\:s\textsuperscript{-1}}, and magnetic fields between 1.0 and 4.0~$\mu$G. Also to establish a basis of comparison, we employ AGN models from \citet{GrovesEtAl2004} with $4Z_\odot$ metallicity, ionization parameters $\log U = - 3$ and $\log U = - 2$ with power-law indices ($F_\nu \propto \nu^{\alpha}$) $\alpha=-1.2$, $\alpha=-1.4$ and $\alpha=-1.7$ that were overplotted in the narrow component. We chose these models because they are the only ones able to explain our data and have been employed in past studies to discern the ionization mechanisms in LINER sources \citep[e.g.][]{RicciEtAl2023}.
    
\begin{figure*}
    \centering
    \includegraphics[width=1.0\textwidth, keepaspectratio]{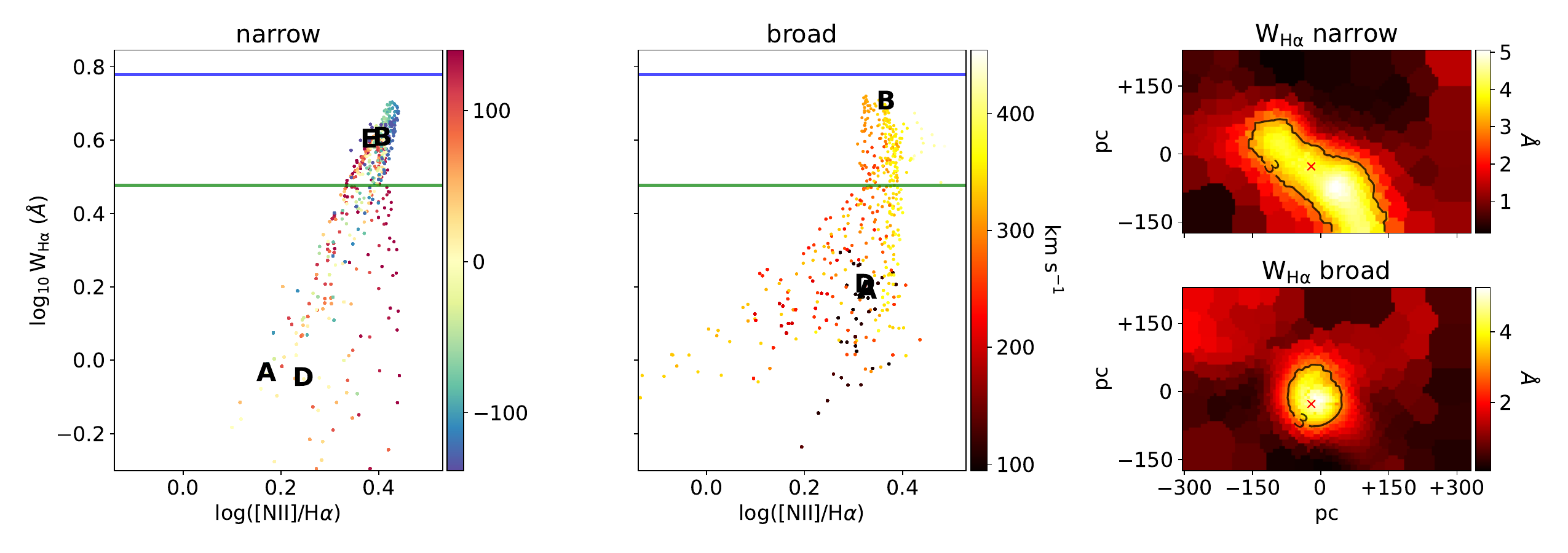}
    \caption{\textbf{Left panel:} WHAN diagram for the narrow component of our data. The points are colour-coded according to the velocity, as in Figs.~{\ref{fig:bpt} and \ref{fig:bpt_oi}}. Different regions in this diagram indicate a prevalence of a given ionization source with AGN showing dominating at $W_{H_{\alpha}}>3$~{\AA}, further diving the region for strong AGNs following \citet{CidFernandesEtAl2011} for $W_{H_{\alpha}}>3$~{\AA}. Moreover, the results obtained for the specific bins defined in Fig.~{\ref{fig:regions}} are overplotted.\textbf{Centre panel:} Same as the left panel but showing the results for the broader component. \textbf{Top-right panel:} the equivalent width of H$\alpha$ for the narrow component, showing where the 3~{\AA} line maps into our data. \textbf{Bottom-right panel:} Same as top-right, but for the broad component.}
    \label{fig:whan}
    \end{figure*}

    In order to make a deeper analysis of the nature of the ionization source of NGC\,6868, we used the WHAN diagram \citep[][]{CidFernandesEtAl2011}. It uses the equivalent width of the H$\alpha$ (W\textsubscript{H$\alpha$}) to measure if the light from hot low-mass evolved stars (HOLMES) is enough to reproduce the ionization levels of the observed emission lines, allowing for the distinction between true AGNs and retired galaxies. Particularly, the 3~\AA\ line in this diagram comes from the bimodal distribution the authors find using SDSS data. This line, therefore, separates galaxies where the dominant ionization source is HOLMES. In Fig.~\ref{fig:whan}, we show the WHAN diagram for each component with the relevant separation lines as discussed. The diagrams show a diverse scenario for the ionization phenomena in NGC\,6868 where the centre falls in the weak AGN region, going to lower and lower values of W\textsubscript{H$\alpha$} moving away from the centre of the galaxy. Looking at the W\textsubscript{H$\alpha$} maps also presented in Fig.~\ref{fig:whan}, in the case of the narrow component, the region we attribute to the central disc is consistent with AGN ionization as well as the central part of the broad component. The contours until 1~{\AA} values in the broad component stretch towards the regions with the peculiar kinematics and flux distributions as described above.

\section{Discussion}

\label{sec:discuss}
    
    \subsection{Detection of an ionized gas disc}
    
\begin{figure}
    \centering
    \includegraphics[width=1.0\columnwidth, keepaspectratio]{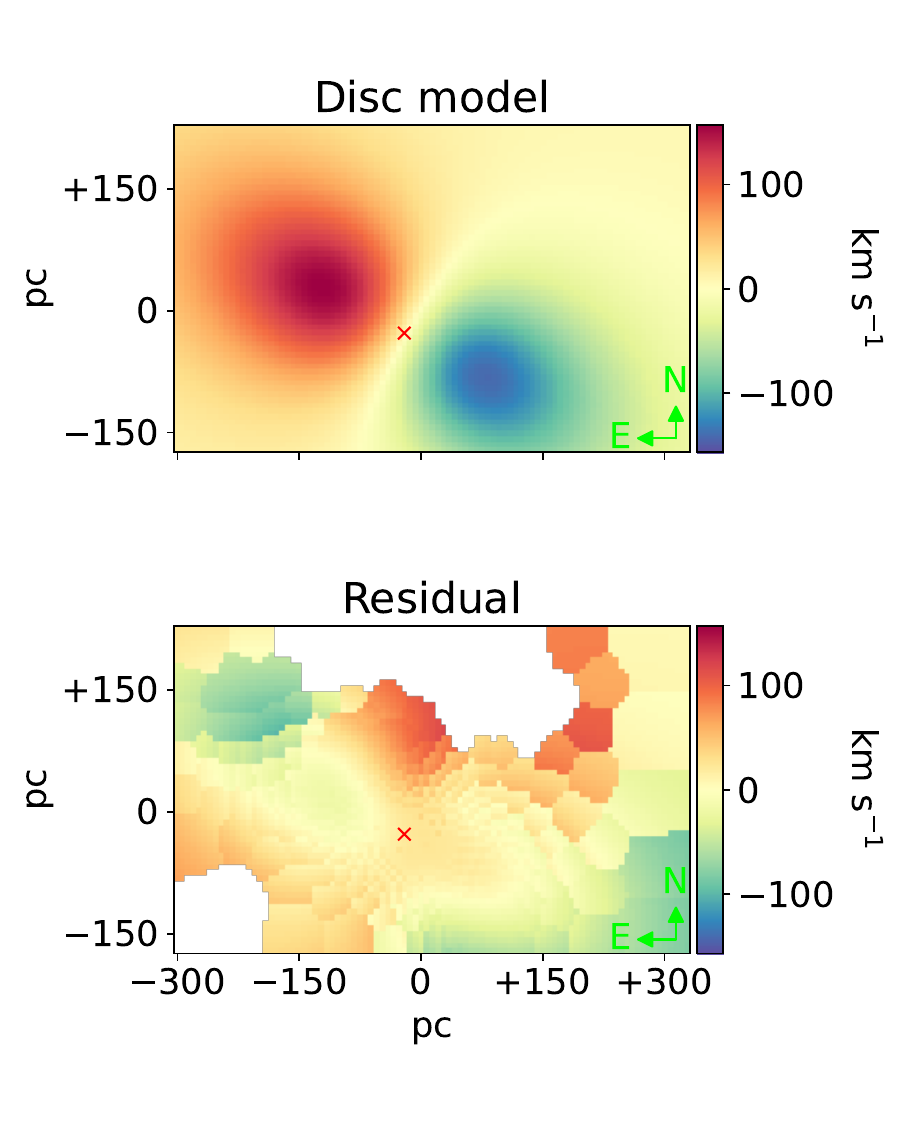}
    \caption{\textbf{Top panel:} Map containing the resulting fit from the kinematical map of the narrow component Fig.~\ref{fig:kinematics}. \textbf{Bottom panel:} residuals (fitted disc model discounted from the observed velocity profile) from the fit. We can confidently model the gas disc as our residuals are kept within \(\sim\)$\pm50$~{km\:s\textsuperscript{-1}} for most of the FoV. White patches are masked regions in the fit due to low S/N.}
    \label{fig:disc_fitting}
    \end{figure}

    The kinematics from the ionized gas derived from our emission line fitting procedure can disentangle the different components that dominate the central region of NGC\,6868. The narrow component that can be seen in Fig.~\ref{fig:kinematics} resembles a rotation profile. Therefore, we used a method within {\sc ifscube} to fit the disc model extracted from \citet[][]{BertolaEtAl1991}. It assumes the gas is orbiting in circular trajectories and follows a velocity field described by
    \begin{equation}
        v_{r}=A\frac{r}{(r^2+c_0^2)^{p/2}}.
        \label{eq:vel}
    \end{equation}
    Thus the projected velocity distribution follows
    \begin{align}
        &v(R, \Psi) = v_{\text{sys}} +\\
        &\frac{A\ R\ \cos{\left(\Psi-\Psi_0\right)}\sin{\Theta}\cos^p{\Theta}}{\left[R^2\left(\sin^2{\left(\Psi-\Psi_0\right)}+\cos^2{\Theta}\cos^2{\left(\Psi-\Psi_0\right)}\right)+c_0^2\cos^2{\Theta}\right]^{p/2} }.\notag
        \label{eq:vel_proj}
    \end{align}
    The resulting fit is shown in Fig.~\ref{fig:disc_fitting} along with the residuals. As can be seen, we can properly model the ionized gas disc with residuals kept within \(\sim\)$\pm50$~{km\:s\textsuperscript{-1}} for most of the FoV. 
    
    This ionized gas disc has already been reported in previous studies \citep[e.g.][]{BusonEtAl1993, ZeilingerEtAl1996, MacchettoEtAl1996,  PizzellaEtAl1997, FerrariEtAl1999, CaonEtAl2000, FerrariEtAl2002}. We observe that the A\textsubscript{V} profile follows roughly the same spatial distribution as the gas disc, as pointed in \citet[][]{HansenEtAl1991}. The molecular gas already reported for NGC\,6868 is found in distances compatible with our findings for the dust lane, therefore the dust is likely shielding the molecular gas from the ionising radiation present in the region. \citet{RoseEtAl2024} investigated the molecular gas kinematics using the CO(2-1) emission line and their results closely match the ones found in this study.

    The main exception to our best model is the region towards the NE (see Fig.~\ref{fig:disc_fitting}) where larger values for the residuals are found to which we provide some hypothesis. The first one is the realisation that the residual profile resembles a rotating disc that could be the remainder of a past merger episode experienced by this galaxy.
    However, we have not found such evidence in the stellar population analysis of this galaxy (Paper~I). We suggest that the more likely scenario involves two main aspects. First, at NE, we see a co-spatial high-velocity redshifted broad component (see Fig.~\ref{fig:kinematics}) which we will discuss the origins of in \S~\ref{origin_emission_profs}. The interaction with this component could have disturbed the gas in the disc producing the observed deviation. Second, as already described, towards the north there is the region where the emission line fluxes drop and are barely detectable. This affects our capability to distinguish both components and, therefore, constrain their kinematics which could have resulted mainly in the redshifted residuals.

    \subsection{What is the origin of the different emission line profiles?}\label{origin_emission_profs}
    
    As can be seen in Fig.~\ref{fig:regions}, one of the characteristics that stand out in NGC\,6868 is the variation in line profiles found. So if the narrow component traces an ionized gas disc, the broad component has a much more diverse nature. 
    
    In the NE region, the broad component traces a high-velocity redshifted component (\(\sim\)$250$~{km\:s\textsuperscript{-1}}). It is also characterised by a similar velocity dispersion when compared to the narrow component ($\sim100$~{km\:s\textsuperscript{-1}}). Beyond the kinematics, line fluxes are enhanced in this direction and to a smaller extent, the A$_V$. 
    
\begin{figure}
    \centering
    \includegraphics[width=1.0\columnwidth, keepaspectratio]{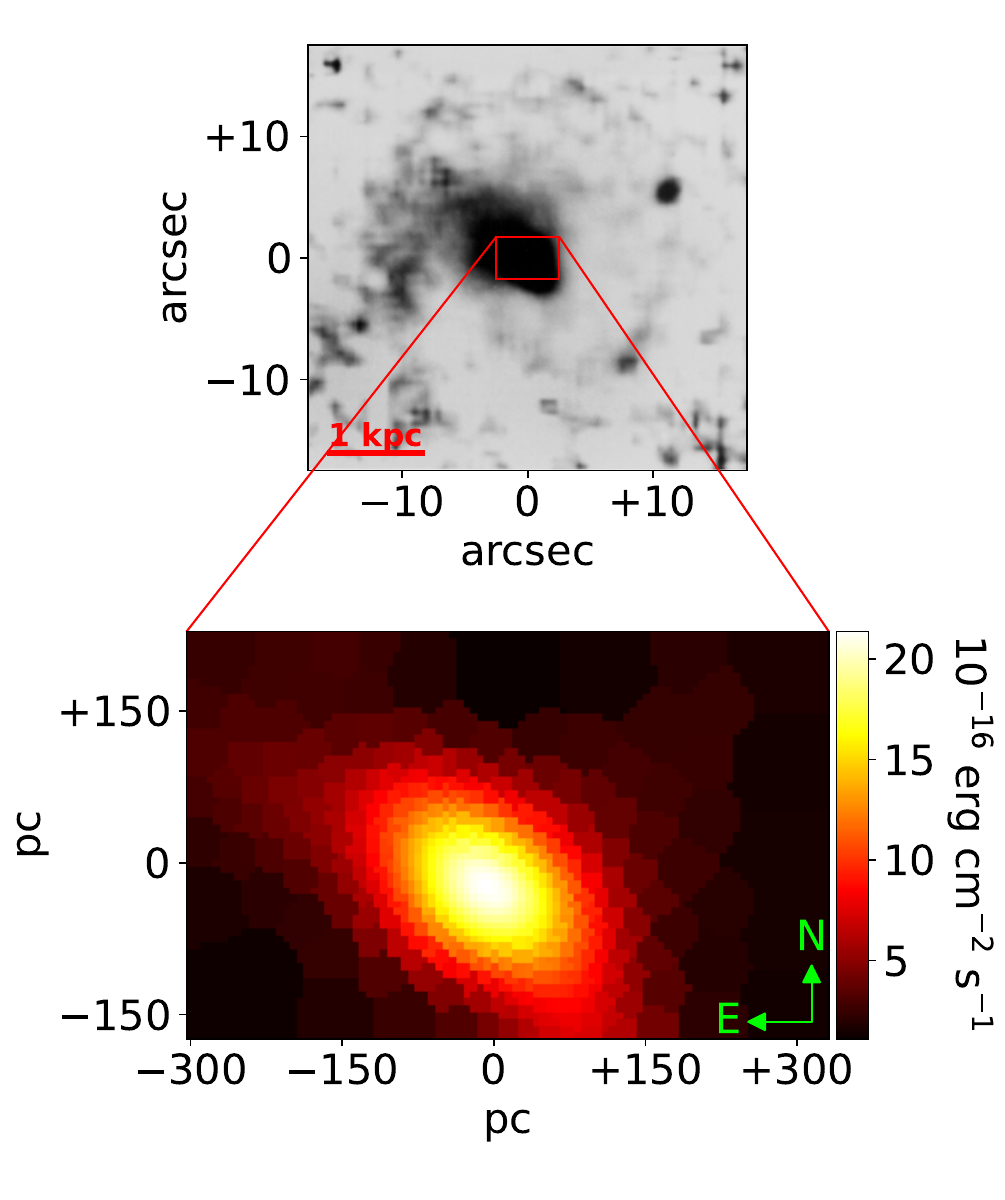}
    \caption{\textbf{Top panel:} Fig.~{2} extracted from \citet{HansenEtAl1991} showing B-r (Gunn r filter) colour for the $35" \times 35"$ inner region of NGC\,6868. Reddened areas are seen in black. Beyond the central dust lane, diffuse elongated components are seen, mainly in the NE direction where a broader structure turning to S is found. \textbf{Bottom panel:} H$\alpha$ flux map for our data cube where the flux towards the NE seems to follow the same direction as the one indicated by \citet{HansenEtAl1991}.}
    \label{fig:hansen_comparison}
    \end{figure}

    \citet[][]{HansenEtAl1991} detected gas filaments with spiral shapes using photometric data in a FoV larger than ours ($35\times35$ arcsec$^2$). In Fig.~\ref{fig:hansen_comparison} we compare \citet[][]{HansenEtAl1991} B-r (Gunn r filter) colour map, which traces the dust in this object, with our H$\alpha$ flux map. From this figure, it is clear that the NE structure we found connects to their description of a spiral arm at NE. Following their interpretation of these spiral structures, this is likely material falling towards the centre of NGC\,6868 forming the spiral structures observed which could come from various sources. It could indeed be a past gas-rich merger or outskirts material travelling towards the galaxy centre. Apparently, this structure has not experienced any significant increase in turbulence and remains kinematically coherent as the velocity dispersion in this region remains similar to the one in the disc. Also, the low electron density values found in this region (Fig.~\ref{fig:dens}) indicate that the inflowing material did not experience any significant compression.
    

    Another of these distinct kinematic components is the blueshifted (\(\sim\)$-100$~{km\:s\textsuperscript{-1}}) broad (\(\sim\)$250$~{km\:s\textsuperscript{-1}}) component (see Fig.~\ref{fig:kinematics}, bottom panels). This region is even clearer in the flux maps (Fig.\ref{fig:fluxes}) where a tail emerges towards the NW as a result of the blue wing found in the line profiles. The higher velocity dispersion hints at a different kinematical origin for this component in which more turbulence has been indicated. It is worth noting that the higher density values towards the centre hit at a gas compression.

    Two scenarios can explain our findings, concerning either an inflow or an outflow from the galaxy centre. The scenario where this traces an inflow is possible because, as we just discussed, this galaxy shows signs of inflowing material towards its centre. Moreover, \citet{RoseEtAl2019} analysing the molecular gas present in this object described the gas as drifting in non-circular orbits, leading to the resulting infall of gas towards the vicinity of the SMBH. In \citet{RoseEtAl2024} they found that the molecular gas is ``predominantly contained in a 0.5 kpc wide and slightly inclined disky structure''. This inflowing material could originate from streams falling towards the centre of the galaxy, and showing different kinematical components (e.g. one moving in the direction of the observer and another away from it), such as those observed in NGC~5044 \citep[a galaxy with similar morphological and physical characteristics as NGC~6868, for details see ][]{DinizEtAl2017}.

    Since NGC\,6868 hosts an LLAGN and presents distinct kinematical characteristics (when compared to the other component at NE) an alternative scenario emerges, e.g. this could be an outflow from the central SMBH. This would also explain the higher velocity dispersion, which indicates that the AGN is inducing turbulence in this gas. A growing number of studies has found that outflows in LINER sources are rather common when their nucleus present AGN signatures \citep[e.g.][]{IlhaEtAl2019, RodriguezdelPinoEtAl2019, RiffelEtAl2019a, Ruschel-DutraEtAl2021, IlhaEtAl2022, HecklerEtAl2022, Deconto-MachadoEtAl2022, HermosaMunozEtAl2022, RiffelEtAl2024, GattoEtAl2024, FalconeEtAl2024, HermosaMunozEtAl2024}, finding outflow mass rates ranging from $10^{-5}$ to 1~{M$_\odot$\:yr$^{-1}$}.
    
\begin{figure*}
    \centering
    \includegraphics[width=1.0\textwidth, keepaspectratio]{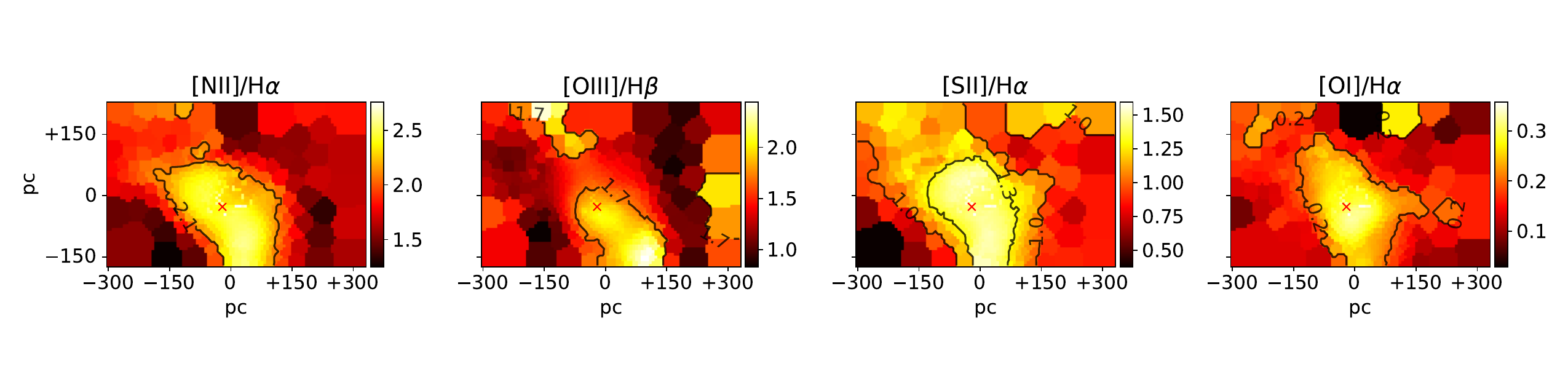}
    \caption{Maps of important emission line ratios. From left to right, [\ion{N}{ii}]/H$\alpha$, [\ion{O}{iii}]/H$\beta$, [\ion{S}{ii}]/H$\alpha$, [\ion{O}{i}]/H$\alpha$. They were all corrected by extinction. }
    \label{fig:ratios}
    \end{figure*}

    In this context, we can estimate the outflow rate considering the following relation:
    \begin{equation}
    \dot{M} = M_\text{out}\frac{v_\text{out}}{r_\text{out}}
    \label{eq:m_dot}
    \end{equation}
    where $\dot{M}$ is the outflow rate, $M_\text{out}$ is the mass of the outflow, $v_\text{out}$ is the velocity of the outflow and $r_\text{out}$ is the spatial extent of the outflow. $M_\text{out}$ can be estimated using Eq.~\ref{eq:ion_mass} applying it only to the spaxels where a component characterised by $v < -95$~{km\:s\textsuperscript{-1}} and $\sigma <260$~{km\:s\textsuperscript{-1}} is detected. For $v_\text{out}$ we assume it to be the absolute median velocity of this component $\sim 110$~{km\:s\textsuperscript{-1}}, and since it reaches the edge of our FoV we use as $r_\text{out}$ the distance from galaxy centre (the red cross in all the figures) to the upper right corner. Of course, we are limited by our FoV so the adopted $r_\text{out}$ is indeed a lower limit. It is also true that other studies have seen that in-situ acceleration could also produce this turbulent component \citep[e.g.]{RodriguezdelPinoEtAl2019, FalconeEtAl2024}. In that case, the radius of the outflow would be simply the size of the region, diminishing the $r_\text{out}$, but making the outflow mass rate get higher (see Eq.~\ref{eq:m_dot}). This results in an outflow mass rate of 0.04~{M$_\odot$\:yr$^{-1}$} which is well within the values found by \citet{HermosaMunozEtAl2024}, further supporting the claim that this emission is due to an outflow from the SMBH. Another evidence supporting this is that the emission line ratios are well explained by models considering shocks. As can be seen in Fig.~{\ref{fig:ratios}}, line ratios seem to be enhanced both in the NE direction towards the edge of the FoV and near the centre in the NW direction, where we detect the peculiar kinematics, indicative of shock-heated gas (see \S~\ref{ionization_driving}).

    
    The region towards the North of our data where no emission lines are detected, also presents a challenging interpretation. We tried to sum more pixels to get a higher S/N to allow us to improve the fit but nearby regions where emission is strong make dealing with the contamination of this rather difficult. Given that the A\textsubscript{V} map does not show any trend towards that region nor in Fig.~\ref{fig:av_gas} nor in the stellar population fits (Paper~I), it is unlikely that this is due to dust extinction. It is more likely that this is a region where devoid of gas. This is further supported by the fact that there is no molecular gas in this same region \citep{RoseEtAl2024}. \citet{MachacekEtAl2010} using X-ray data, found X-ray cavities in larger scales which they have correlated with an echo of AGN activity. 
    

    In the SW region, we only need one Gaussian component to fit the data. At first glance, this seems to only be a part of the ionised gas disc which, kinematically is true, but when we look at the flux of [\ion{O}{iii}]~$\lambda5007$ it is evident that the SW shows a peculiar behaviour also maintained in the [\ion{O}{iii}]~$\lambda5007$/H$\beta$ map with a clear enhancement of the [\ion{O}{iii}]~$\lambda5007$ flux. Because this is a high ionization ion some form of added excitation mechanism is necessary to explain this observation. Given that the temperature increases towards this region (Fig.~\ref{fig:temp}) and that \citet{RoseEtAl2024} has detected a stream of molecular gas that extends beyond our FoV, we suggest that this enhancement is due to the interaction with this stream. 

    \subsection{What is driving the gas ionization in NGC 6868?}\label{ionization_driving}
    
    NGC\,6868 displays an extended ionized gas component as can be seen in Fig.~\ref{fig:big_picture} being classified as a LINER object in the BPT diagram. The ionization mechanisms behind these objects are diverse and IFS studies have allowed for the disentanglement of them. In this sense, NGC\,6868 seems to have a huge variety of excitation mechanisms.
    
    It is already known that this galaxy hosts an LLAGN due to the detection of a radio source, an X-ray core as well as [\ion{Ne}{v}] \citep{SheEtAl2017, BiEtAl2020, RampazzoEtAl2013, RicciEtAl2023}, despite the non-detection of a broad component in \ion{H}{i} lines. From the WHAN diagram (Fig.~\ref{fig:whan}), we see that the central regions in both the narrow and broad components require the energetic input of an AGN which is consistent with the LLAGN picture, also coinciding with AGN models in BPT diagram of $\alpha=-1.2,-1.4$ and $\log U=-3, -2$ (Fig.~\ref{fig:bpt_oi}).
    
    Looking at the narrow-line BPT diagrams we see that [\ion{O}{iii}]~$\lambda5007$/H$\beta$ always present a lower value in the centre compared to the disc, also the clear gradient in the temperature profile where temperature increases outwards from the centre are all good evidence that shocks have some importance in the ionization balance for this object.

    [\ion{O}{i}]~$\lambda6300$ is a famous tracer for shocks and as can be seen in Fig.~\ref{fig:bpt_oi} shock models can explain the emission line ratios found. \citet{HoEtAl2014} and \citet{AllenEtAl2008} argument that for $\sigma > 150$~{km\:s\textsuperscript{-1}} and [\ion{O}{i}]~$\lambda6300$/H$\alpha > -1.0$ would be enough to say that shocks are the dominant mechanisms. In this reasoning, we show in the bottom panel of Fig.~\ref{fig:bpt_oi} that the line ratios are consistent with shock models with magnetic field values between 2 and 4~{$\mu G$} and an average velocity of 350~{km\:s\textsuperscript{-1}}. This coupled with the fact that we see the outward increasing temperature profile, surpassing the 20000~{K} in some points provides even more evidence that the blue-shifted component is an actual outflow that is shock-heating the surrounding medium. Additionally, the [\ion{S}{ii}]~$\lambda6716,6731$/H$\alpha$ diagram also shows that the line ratio increases towards the redshifted and the blueshifted components ( e.g. comparing the positions of regions B, D and A in the diagrams it becomes clear). By comparing the narrow and broad component results, the gradient inverts, again providing more evidence that another source of excitation must be present which we interpret as shocks. It is worth mentioning that we cannot say if shocks are the dominant source of ionization, but surely they contribute to the ionization balance of this object.
    
    
    Due to the ubiquitous old stellar population, it is expected that HOLMES create a dispersed ionizing radiation field capable of ionizing the surrounding gas. More evidence of the presence of this ionization source is the WHAN diagram and the $W_{H_{\alpha}}$ maps. Looking at the regions where $W_{H_{\alpha}}<1$~{\AA} the ionization is likely to be provided by HOLMES. In the region $W_{H_{\alpha}}>3$~{\AA} the most likely scenario is AGN ionization which is compatible with all the other diagrams. Looking at the spatial distribution of $W_{H_{\alpha}}$ it is clear that it is between 1 and 3~{\AA} where all the regions with distinct kinematical features coincide, contributing to the picture that, despite playing a secondary role, when combined with the kinematical and temperature findings, we conclude that shocks also contribute to the gas excitation.


    Finally, recent studies \citep[e.g.][]{LagosEtAl2022, RicciEtAl2023, HermosaMunozEtAl2024, HsiehEtAl2017} have shown that the physical processes dominating the gas ionization, in the central region of sources classified as LINERs, are due to LLAGNs, while the of nuclear component is the one with divergent results. For instance, \citet{LagosEtAl2022} prefers the scenario of pAGB stars ionizing the extended component, whereas \citet{RicciEtAl2023} states the importance of shocks for some of their sources. In this context, NGC\,6868 proves to be a galaxy that shares characteristics with both studies, showing that most likely both mechanisms are dominating in different regions. 

    \color{black}

\section{Conclusions}
\label{sec:conclusion}

    We analysed the central region of NGC\,6868 using GMOS-IFU and mapped the properties of the ionized gas as well as the excitation mechanism behind the gas ionization. The complexity in the gas kinematics and ionization makes NGC\,6868 an exciting laboratory to study the different mechanisms involved in powering LINER sources. Our main findings are summarised as flows:
    
    \begin{itemize}
        \item Channel maps and line profiles reveal complex kinematics and morphology, hinting at different processes acting on NGC\,6868.
        \item Emission line fitting has revealed two kinematic components: a narrow and a broad.
        \item The narrow component traces an ionized gas disc and the broad component traces inflowing gas that is being driven to the vicinity of the LLAGN present in NGC\,6868, settling in a dispersion-dominated component. Flux distributions follow these findings.
        \item The spatial distribution of the reddening obtained from Balmer decrement is consistent with the derived from the stellar continuum fitting albeit with a larger amplitude. This is likely due to the obscuration of the ionizing source, mainly in the central regions.
        \item We report the first measurement of electron temperature in NGC\,6868. We find a temperature of \(\sim\)$14000$~{K} for the central region with the outer parts surpassing the \(\sim\)$20000$~{K}. The density profile shows an inverse behaviour as the central regions show an enhancement (\(\sim\)$800$~{cm\textsuperscript{-3}}) when compared to the outer regions(\(\sim\)$100$~{cm\textsuperscript{-3}}), hinting at a compression of the ionized gas.
        \item All the points from our observations are found within the LINER region in the BPT diagram. Coupled with the WHAN diagram, we conclude that the central region is ionized mainly by an LLAGN, while HOLMES create an ionizing radiation field that is responsible for the ionization at larger scales.
        \item The derived temperature profile, coupled with the kinematics and emission line ratios point towards the presence of shock-heated gas.
    \end{itemize}
    
    Our work reinforces the necessity of a detailed analysis via IFU of LINERs to disentangle the different mechanisms that drive the observed emission. For instance, the presence of an outflow component in LINERs arises now as an important ingredient to explain the observed spectrum of these objects.

\color{black}
\section*{Acknowledgements}
    The authors thank Jose Andres Hernandez Jimenez for helpful discussions. We thank the anonymous referee for insightful comments and discussion. This work was supported by Brazilian funding agencies Conselho Nacional de Desenvolvimento Cient\'{i}fico e Tecnol\'ogico (CNPq) and Coordena\c{c}\~ao de Aperfei\c{c}oamento de Pessoal de N\'{i}vel Superior (CAPES) and by the \textit{Programa de Pós-Graduação em Física} (PPGFis) at UFRGS. JPVB acknowledges financial support from CNPq and CAPES (Proj. 0001) and the support of a fellowship from the ”la Caixa” Foundation (ID 100010434). The fellowship code is LCF/BQ/DI23/11990084. RR thanks CNPq (Proj. 311223/2020-6, 304927/2017-1 and 400352/2016-8), Funda\c{c}\~ao de amparo \`{a} pesquisa do Rio Grande do Sul (FAPERGS, Proj. 16/2551-0000251-7 and 19/1750-2) and CAPES (Proj. 0001). TVR thanks CNPq for support under grant 306790/2019-0 and 304584/2022-3. RAR acknowledges the support from CNPq (Proj. 303450/2022-3, 403398/2023-1, \& 441722/2023-7), FAPERGS (Proj. 21/2551-0002018-0), and CAPES (Proj. 88887.894973/2023-00). MT thanks the support of CNPq (process 312541/2021-0) and the program L’Oréal UNESCO ABC \textit{Para Mulheres na Ciência}. LGDH acknowledges support by National Key R\&D Program of China No.2022YFF0503402 and National Natural Science Foundation of China (NSFC) project number E345251001. DRD acknowledges financial support from CNPq (Proj. 313040/2022-2). AFM has received support from PID2021-123313NA-I00 and 4RYC2021-031099-I of the MICIN/AEI/10.13039/501100011033/UE.

    Based on observations obtained at the international Gemini Observatory and processed using the Gemini {\sc iraf} package, a program of NSF’s NOIRLab, which is managed by the Association of Universities for Research in Astronomy (AURA) under a cooperative agreement with the National Science Foundation on behalf of the Gemini Observatory partnership: the National Science Foundation (United States), National Research Council (Canada), Agencia Nacional de Investigación y Desarrollo (Chile), Ministerio de Ciencia, Tecnología e Innovación (Argentina), Ministério da Ciência, Tecnologia, Inovações e Comunicações (Brazil), and Korea Astronomy and Space Science Institute (Republic of Korea).

\section*{Data Availability}

The data are publicly available on {\sc gemini} archive under the project GS-2013A-Q-52. The reduced and processed data can be made available under reasonable request.



\bibliographystyle{mnras}
\bibliography{references} 

\begin{thebibliography}{}
\makeatletter
\relax
\def\mn@urlcharsother{\let\do\@makeother \do\$\do\&\do\#\do\^\do\_\do\%\do\~}
\def\mn@doi{\begingroup\mn@urlcharsother \@ifnextchar [ {\mn@doi@}
  {\mn@doi@[]}}
\def\mn@doi@[#1]#2{\def\@tempa{#1}\ifx\@tempa\@empty \href
  {http://dx.doi.org/#2} {doi:#2}\else \href {http://dx.doi.org/#2} {#1}\fi
  \endgroup}
\def\mn@eprint#1#2{\mn@eprint@#1:#2::\@nil}
\def\mn@eprint@arXiv#1{\href {http://arxiv.org/abs/#1} {{\tt arXiv:#1}}}
\def\mn@eprint@dblp#1{\href {http://dblp.uni-trier.de/rec/bibtex/#1.xml}
  {dblp:#1}}
\def\mn@eprint@#1:#2:#3:#4\@nil{\def\@tempa {#1}\def\@tempb {#2}\def\@tempc
  {#3}\ifx \@tempc \@empty \let \@tempc \@tempb \let \@tempb \@tempa \fi \ifx
  \@tempb \@empty \def\@tempb {arXiv}\fi \@ifundefined
  {mn@eprint@\@tempb}{\@tempb:\@tempc}{\expandafter \expandafter \csname
  mn@eprint@\@tempb\endcsname \expandafter{\@tempc}}}

\bibitem[\protect\citeauthoryear{Allen, Groves, Dopita, Sutherland  \&
  Kewley}{Allen et~al.}{2008}]{AllenEtAl2008}
Allen M.~G.,  Groves B.~A.,  Dopita M.~A.,  Sutherland R.~S.,   Kewley L.~J.,
  2008, \mn@doi [The Astrophysical Journal Supplement Series] {10.1086/589652},
  178, 20

\bibitem[\protect\citeauthoryear{Babyk, McNamara, Nulsen, Hogan, Vantyghem,
  Russell, Pulido  \& Edge}{Babyk et~al.}{2018}]{BabykEtAl2018}
Babyk I.~V.,  McNamara B.~R.,  Nulsen P. E.~J.,  Hogan M.~T.,  Vantyghem A.~N.,
   Russell H.~R.,  Pulido F.~A.,   Edge A.~C.,  2018, \mn@doi [The
  Astrophysical Journal] {10.3847/1538-4357/aab3c9}, 857, 32

\bibitem[\protect\citeauthoryear{Baldwin, Phillips  \& Terlevich}{Baldwin
  et~al.}{1981}]{BaldwinEtAl1981}
Baldwin J.~A.,  Phillips M.~M.,   Terlevich R.,  1981, \mn@doi [Publications of
  the Astronomical Society of the Pacific] {10.1086/130766}, 93, 5

\bibitem[\protect\citeauthoryear{Barth \& Shields}{Barth \&
  Shields}{2000}]{BarthShields2000}
Barth A.~J.,  Shields J.~C.,  2000, \mn@doi [Publications of the Astronomical
  Society of the Pacific] {10.1086/316583}, 112, 753

\bibitem[\protect\citeauthoryear{Belfiore et~al.,}{Belfiore
  et~al.}{2015}]{BelfioreEtAl2015}
Belfiore F.,  et~al., 2015, \mn@doi [Monthly Notices of the Royal Astronomical
  Society] {10.1093/mnras/stv296}, 449, 867

\bibitem[\protect\citeauthoryear{Belfiore et~al.,}{Belfiore
  et~al.}{2016}]{BelfioreEtAl2016}
Belfiore F.,  et~al., 2016, \mn@doi [Monthly Notices of the Royal Astronomical
  Society] {10.1093/mnras/stw1234}, 461, 3111

\bibitem[\protect\citeauthoryear{Benedetti et~al.,}{Benedetti
  et~al.}{2023}]{BenedettiEtAl2023}
Benedetti J. P.~V.,  et~al., 2023, \mn@doi [Monthly Notices of the Royal
  Astronomical Society] {10.1093/mnras/stad1148}, 522, 2570

\bibitem[\protect\citeauthoryear{Bertola, Bettoni, Danziger, Sadler, Sparke  \&
  de Zeeuw}{Bertola et~al.}{1991}]{BertolaEtAl1991}
Bertola F.,  Bettoni D.,  Danziger J.,  Sadler E.,  Sparke L.,   de Zeeuw T.,
  1991, \mn@doi [The Astrophysical Journal] {10.1086/170058}, 373, 369

\bibitem[\protect\citeauthoryear{Bi, Feng  \& Ho}{Bi et~al.}{2020}]{BiEtAl2020}
Bi S.,  Feng H.,   Ho L.~C.,  2020, \mn@doi [The Astrophysical Journal]
  {10.3847/1538-4357/aba761}, 900, 124

\bibitem[\protect\citeauthoryear{Binette, Magris, Stasińska  \&
  Bruzual}{Binette et~al.}{1994}]{BinetteEtAl1994}
Binette L.,  Magris C.~G.,  Stasińska G.,   Bruzual A.~G.,  1994, Astronomy
  and Astrophysics, 292, 13

\bibitem[\protect\citeauthoryear{Bonatto, Bica  \& Alloin}{Bonatto
  et~al.}{1989}]{BonattoEtAl1989}
Bonatto C.,  Bica E.,   Alloin D.,  1989, Astronomy and Astrophysics, 226, 23

\bibitem[\protect\citeauthoryear{Bregman, Snider, Grego  \& Cox}{Bregman
  et~al.}{1998}]{BregmanEtAl1998}
Bregman J.~N.,  Snider B.~A.,  Grego L.,   Cox C.~V.,  1998, \mn@doi [The
  Astrophysical Journal] {10.1086/305653}, 499, 670

\bibitem[\protect\citeauthoryear{Buson et~al.,}{Buson
  et~al.}{1993}]{BusonEtAl1993}
Buson L.~M.,  et~al., 1993, Astronomy and Astrophysics, 280, 409

\bibitem[\protect\citeauthoryear{Caon, Macchetto  \& Pastoriza}{Caon
  et~al.}{2000}]{CaonEtAl2000}
Caon N.,  Macchetto D.,   Pastoriza M.,  2000, \mn@doi [The Astrophysical
  Journal Supplement Series] {10.1086/313315}, 127, 39

\bibitem[\protect\citeauthoryear{Cappellari \& Copin}{Cappellari \&
  Copin}{2003}]{CappellariCopin2003}
Cappellari M.,  Copin Y.,  2003, \mn@doi [Monthly Notices of the Royal
  Astronomical Society] {10.1046/j.1365-8711.2003.06541.x}, 342, 345

\bibitem[\protect\citeauthoryear{Cardelli, Clayton  \& Mathis}{Cardelli
  et~al.}{1989}]{CardelliEtAl1989}
Cardelli J.~A.,  Clayton G.~C.,   Mathis J.~S.,  1989, \mn@doi [The
  Astrophysical Journal] {10.1086/167900}, 345, 245

\bibitem[\protect\citeauthoryear{Carrasco, Oliveira  \& Infante}{Carrasco
  et~al.}{2006}]{CarrascoEtAl2006}
Carrasco E.~R.,  Oliveira C. M.~d.,   Infante L.,  2006, \mn@doi [The
  Astronomical Journal] {10.1086/507447}, 132, 1796

\bibitem[\protect\citeauthoryear{Cid~Fernandes, Mateus, Sodré, Stasińska  \&
  Gomes}{Cid~Fernandes et~al.}{2005}]{CidFernandesEtAl2005}
Cid~Fernandes R.,  Mateus A.,  Sodré L.,  Stasińska G.,   Gomes J.~M.,  2005,
  \mn@doi [Monthly Notices of the Royal Astronomical Society]
  {10.1111/j.1365-2966.2005.08752.x}, 358, 363

\bibitem[\protect\citeauthoryear{Cid~Fernandes, Stasińska, Mateus  \&
  Vale~Asari}{Cid~Fernandes et~al.}{2011}]{CidFernandesEtAl2011}
Cid~Fernandes R.,  Stasińska G.,  Mateus A.,   Vale~Asari N.,  2011, \mn@doi
  [Monthly Notices of the Royal Astronomical Society]
  {10.1111/j.1365-2966.2011.18244.x}, 413, 1687

\bibitem[\protect\citeauthoryear{Constantin \& Vogeley}{Constantin \&
  Vogeley}{2006}]{ConstantinVogeley2006}
Constantin A.,  Vogeley M.~S.,  2006, \mn@doi [The Astrophysical Journal]
  {10.1086/507087}, 650, 727

\bibitem[\protect\citeauthoryear{Croton et~al.,}{Croton
  et~al.}{2006}]{CrotonEtAl2006}
Croton D.~J.,  et~al., 2006, \mn@doi [Monthly Notices of the Royal Astronomical
  Society] {10.1111/j.1365-2966.2005.09675.x}, 365, 11

\bibitem[\protect\citeauthoryear{Deconto-Machado et~al.,}{Deconto-Machado
  et~al.}{2022}]{Deconto-MachadoEtAl2022}
Deconto-Machado A.,  et~al., 2022, \mn@doi [Astronomy \& Astrophysics]
  {10.1051/0004-6361/202140613}, 659, A131

\bibitem[\protect\citeauthoryear{Diniz, Pastoriza, Hernandez-Jimenez, Riffel,
  Ricci, Steiner  \& Riffel}{Diniz et~al.}{2017}]{DinizEtAl2017}
Diniz S. I.~F.,  Pastoriza M.~G.,  Hernandez-Jimenez J.~A.,  Riffel R.,  Ricci
  T.~V.,  Steiner J.~E.,   Riffel R.~A.,  2017, \mn@doi [Monthly Notices of the
  Royal Astronomical Society] {10.1093/mnras/stx1322}, 470, 1703

\bibitem[\protect\citeauthoryear{Dopita \& Sutherland}{Dopita \&
  Sutherland}{1995}]{DopitaSutherland1995}
Dopita M.~A.,  Sutherland R.~S.,  1995, \mn@doi [The Astrophysical Journal]
  {10.1086/176596}, 455, 468

\bibitem[\protect\citeauthoryear{Eracleous, Hwang  \& Flohic}{Eracleous
  et~al.}{2010}]{EracleousEtAl2010}
Eracleous M.,  Hwang J.~A.,   Flohic H. M. L.~G.,  2010, \mn@doi [The
  Astrophysical Journal] {10.1088/0004-637X/711/2/796}, 711, 796

\bibitem[\protect\citeauthoryear{Fabian}{Fabian}{2012}]{Fabian2012}
Fabian A.,  2012, \mn@doi [Annual Review of Astronomy and Astrophysics]
  {10.1146/annurev-astro-081811-125521}, 50, 455

\bibitem[\protect\citeauthoryear{Falcone et~al.,}{Falcone
  et~al.}{2024}]{FalconeEtAl2024}
Falcone J.,  et~al., 2024, \mn@doi [The Astrophysical Journal]
  {10.3847/1538-4357/ad5283}, 971, 17

\bibitem[\protect\citeauthoryear{Ferland \& Netzer}{Ferland \&
  Netzer}{1983}]{FerlandNetzer1983}
Ferland G.~J.,  Netzer H.,  1983, \mn@doi [The Astrophysical Journal]
  {10.1086/160577}, 264, 105

\bibitem[\protect\citeauthoryear{Ferrari, Pastoriza, Macchetto  \&
  Caon}{Ferrari et~al.}{1999}]{FerrariEtAl1999}
Ferrari F.,  Pastoriza M.~G.,  Macchetto F.,   Caon N.,  1999, \mn@doi
  [Astronomy and Astrophysics Supplement Series] {10.1051/aas:1999465}, 136,
  269

\bibitem[\protect\citeauthoryear{Ferrari, Pastoriza, Macchetto, Bonatto,
  Panagia  \& Sparks}{Ferrari et~al.}{2002}]{FerrariEtAl2002}
Ferrari F.,  Pastoriza M.~G.,  Macchetto F.~D.,  Bonatto C.,  Panagia N.,
  Sparks W.~B.,  2002, \mn@doi [Astronomy \& Astrophysics]
  {10.1051/0004-6361:20020582}, 389, 355

\bibitem[\protect\citeauthoryear{Flohic, Eracleous, Chartas, Shields  \&
  Moran}{Flohic et~al.}{2006}]{FlohicEtAl2006}
Flohic H. M. L.~G.,  Eracleous M.,  Chartas G.,  Shields J.~C.,   Moran E.~C.,
  2006, \mn@doi [The Astrophysical Journal] {10.1086/505296}, 647, 140

\bibitem[\protect\citeauthoryear{Gatto, Storchi-Bergmann, Riffel, Riffel,
  Rembold, Schimoia, Mallmann  \& Ilha}{Gatto et~al.}{2024}]{GattoEtAl2024}
Gatto L.,  Storchi-Bergmann T.,  Riffel R.~A.,  Riffel R.,  Rembold S.~B.,
  Schimoia J.~S.,  Mallmann N.~D.,   Ilha G.~S.,  2024, \mn@doi [Monthly
  Notices of the Royal Astronomical Society] {10.1093/mnras/stae989}, 530, 3059

\bibitem[\protect\citeauthoryear{Gebhardt et~al.,}{Gebhardt
  et~al.}{2000}]{GebhardtEtAl2000}
Gebhardt K.,  et~al., 2000, \mn@doi [The Astrophysical Journal]
  {10.1086/312840}, 539, L13

\bibitem[\protect\citeauthoryear{Girardi, Bressan, Bertelli  \& Chiosi}{Girardi
  et~al.}{2000}]{GirardiEtAl2000}
Girardi L.,  Bressan A.,  Bertelli G.,   Chiosi C.,  2000, \mn@doi [Astronomy
  and Astrophysics Supplement Series] {10.1051/aas:2000126}, 141, 371

\bibitem[\protect\citeauthoryear{Gomes et~al.,}{Gomes
  et~al.}{2016}]{GomesEtAl2016}
Gomes J.~M.,  et~al., 2016, \mn@doi [Astronomy \& Astrophysics]
  {10.1051/0004-6361/201525976}, 588, A68

\bibitem[\protect\citeauthoryear{Gonzalez \& Woods}{Gonzalez \&
  Woods}{2008}]{GonzalezWoods2008}
Gonzalez R.~C.,  Woods R.~E.,  2008, Digital image processing, 3rd ed edn.
Prentice Hall, Upper Saddle River, N.J

\bibitem[\protect\citeauthoryear{Groves, Dopita  \& Sutherland}{Groves
  et~al.}{2004}]{GrovesEtAl2004}
Groves B.~A.,  Dopita M.~A.,   Sutherland R.~S.,  2004, \mn@doi [The
  Astrophysical Journal Supplement Series] {10.1086/421113}, 153, 9

\bibitem[\protect\citeauthoryear{Halpern \& Steiner}{Halpern \&
  Steiner}{1983}]{HalpernSteiner1983}
Halpern J.~P.,  Steiner J.~E.,  1983, \mn@doi [The Astrophysical Journal]
  {10.1086/184051}, 269, L37

\bibitem[\protect\citeauthoryear{Hamuy, Walker, Suntzeff, Gigoux, Heathcote  \&
  Phillips}{Hamuy et~al.}{1992}]{HamuyEtAl1992}
Hamuy M.,  Walker A.~R.,  Suntzeff N.~B.,  Gigoux P.,  Heathcote S.~R.,
  Phillips M.~M.,  1992, \mn@doi [Publications of the Astronomical Society of
  the Pacific] {10.1086/133028}, 104, 533

\bibitem[\protect\citeauthoryear{Hansen, Jorgensen  \& Norgaard-Nielsen}{Hansen
  et~al.}{1991}]{HansenEtAl1991}
Hansen L.,  Jorgensen H.~E.,   Norgaard-Nielsen H.~U.,  1991, Astronomy and
  Astrophysics, 243, 49

\bibitem[\protect\citeauthoryear{Healey, Romani, Taylor, Sadler, Ricci, Murphy,
  Ulvestad  \& Winn}{Healey et~al.}{2007}]{HealeyEtAl2007}
Healey S.~E.,  Romani R.~W.,  Taylor G.~B.,  Sadler E.~M.,  Ricci R.,  Murphy
  T.,  Ulvestad J.~S.,   Winn J.~N.,  2007, \mn@doi [The Astrophysical Journal
  Supplement Series] {10.1086/513742}, 171, 61

\bibitem[\protect\citeauthoryear{Heckler, Ricci  \& Riffel}{Heckler
  et~al.}{2022}]{HecklerEtAl2022}
Heckler K.~F.,  Ricci T.~V.,   Riffel R.~A.,  2022, \mn@doi [Monthly Notices of
  the Royal Astronomical Society] {10.1093/mnras/stac3041}, 517, 5959

\bibitem[\protect\citeauthoryear{Heckman}{Heckman}{1980}]{Heckman1980}
Heckman T.~M.,  1980, Astronomy and Astrophysics, 87, 152

\bibitem[\protect\citeauthoryear{Heckman \& Best}{Heckman \&
  Best}{2014}]{HeckmanBest2014}
Heckman T.~M.,  Best P.~N.,  2014, \mn@doi [Annual Review of Astronomy and
  Astrophysics] {10.1146/annurev-astro-081913-035722}, 52, 589

\bibitem[\protect\citeauthoryear{Hermosa~Muñoz, Márquez, Cazzoli, Masegosa
  \& Agís-González}{Hermosa~Muñoz et~al.}{2022}]{HermosaMunozEtAl2022}
Hermosa~Muñoz L.,  Márquez I.,  Cazzoli S.,  Masegosa J.,   Agís-González
  B.,  2022, \mn@doi [Astronomy \& Astrophysics] {10.1051/0004-6361/202142629},
  660, A133

\bibitem[\protect\citeauthoryear{Hermosa~Muñoz et~al.,}{Hermosa~Muñoz
  et~al.}{2024}]{HermosaMunozEtAl2024}
Hermosa~Muñoz L.,  et~al., 2024, \mn@doi [Astronomy \& Astrophysics]
  {10.1051/0004-6361/202347675}, 683, A43

\bibitem[\protect\citeauthoryear{Herpich, Mateus, Stasińska, Cid Fernandes
  \& Vale Asari}{Herpich et~al.}{2016}]{HerpichEtAl2016}
Herpich F.,  Mateus A.,  Stasińska G.,  Cid Fernandes R.,   Vale Asari N.,
  2016, \mn@doi [Monthly Notices of the Royal Astronomical Society]
  {10.1093/mnras/stw1742}, 462, 1826

\bibitem[\protect\citeauthoryear{Ho}{Ho}{2008}]{Ho2008}
Ho L.~C.,  2008, \mn@doi [Annual Review of Astronomy and Astrophysics]
  {10.1146/annurev.astro.45.051806.110546}, 46, 475

\bibitem[\protect\citeauthoryear{Ho, Filippenko  \& Sargent}{Ho
  et~al.}{1993}]{HoEtAl1993}
Ho L.~C.,  Filippenko A.~V.,   Sargent W. L.~W.,  1993, \mn@doi [The
  Astrophysical Journal] {10.1086/173291}, 417, 63

\bibitem[\protect\citeauthoryear{Ho, Filippenko  \& Sargent}{Ho
  et~al.}{1996}]{HoEtAl1996}
Ho L.~C.,  Filippenko A.~V.,   Sargent W. L.~W.,  1996, \mn@doi [The
  Astrophysical Journal] {10.1086/177140}, 462, 183

\bibitem[\protect\citeauthoryear{Ho, Filippenko, Sargent  \& Peng}{Ho
  et~al.}{1997}]{HoEtAl1997}
Ho L.~C.,  Filippenko A.~V.,  Sargent W. L.~W.,   Peng C.~Y.,  1997, \mn@doi
  [The Astrophysical Journal Supplement Series] {10.1086/313042}, 112, 391

\bibitem[\protect\citeauthoryear{Ho et~al.,}{Ho et~al.}{2014}]{HoEtAl2014}
Ho I.-T.,  et~al., 2014, \mn@doi [Monthly Notices of the Royal Astronomical
  Society] {10.1093/mnras/stu1653}, 444, 3894

\bibitem[\protect\citeauthoryear{Ho et~al.,}{Ho et~al.}{2016}]{HoEtAl2016}
Ho I.-T.,  et~al., 2016, \mn@doi [Monthly Notices of the Royal Astronomical
  Society] {10.1093/mnras/stw017}, 457, 1257

\bibitem[\protect\citeauthoryear{Hsieh et~al.,}{Hsieh
  et~al.}{2017}]{HsiehEtAl2017}
Hsieh B.~C.,  et~al., 2017, \mn@doi [The Astrophysical Journal]
  {10.3847/2041-8213/aa9d80}, 851, L24

\bibitem[\protect\citeauthoryear{Häring \& Rix}{Häring \&
  Rix}{2004}]{HaringRix2004}
Häring N.,  Rix H.-W.,  2004, \mn@doi [The Astrophysical Journal]
  {10.1086/383567}, 604, L89

\bibitem[\protect\citeauthoryear{Ilha et~al.,}{Ilha
  et~al.}{2019}]{IlhaEtAl2019}
Ilha G.~S.,  et~al., 2019, \mn@doi [Monthly Notices of the Royal Astronomical
  Society] {10.1093/mnras/sty3373}, 484, 252

\bibitem[\protect\citeauthoryear{Ilha et~al.,}{Ilha
  et~al.}{2022}]{IlhaEtAl2022}
Ilha G.~S.,  et~al., 2022, \mn@doi [Monthly Notices of the Royal Astronomical
  Society] {10.1093/mnras/stac2233}, 516, 1442

\bibitem[\protect\citeauthoryear{Kauffmann et~al.,}{Kauffmann
  et~al.}{2003}]{KauffmannEtAl2003}
Kauffmann G.,  et~al., 2003, \mn@doi [Monthly Notices of the Royal Astronomical
  Society] {10.1046/j.1365-8711.2003.06292.x}, 341, 54

\bibitem[\protect\citeauthoryear{Kehrig et~al.,}{Kehrig
  et~al.}{2012}]{KehrigEtAl2012}
Kehrig C.,  et~al., 2012, \mn@doi [Astronomy \& Astrophysics]
  {10.1051/0004-6361/201118357}, 540, A11

\bibitem[\protect\citeauthoryear{Kewley, Groves, Kauffmann  \& Heckman}{Kewley
  et~al.}{2006}]{KewleyEtAl2006}
Kewley L.~J.,  Groves B.,  Kauffmann G.,   Heckman T.,  2006, \mn@doi [Monthly
  Notices of the Royal Astronomical Society]
  {10.1111/j.1365-2966.2006.10859.x}, 372, 961

\bibitem[\protect\citeauthoryear{Kormendy \& Ho}{Kormendy \&
  Ho}{2013}]{KormendyHo2013}
Kormendy J.,  Ho L.~C.,  2013, \mn@doi [Annual Review of Astronomy and
  Astrophysics] {10.1146/annurev-astro-082708-101811}, 51, 511

\bibitem[\protect\citeauthoryear{Lagos et~al.,}{Lagos
  et~al.}{2022}]{LagosEtAl2022}
Lagos P.,  et~al., 2022, \mn@doi [Monthly Notices of the Royal Astronomical
  Society] {10.1093/mnras/stac2535}

\bibitem[\protect\citeauthoryear{Lauberts \& Valentijn}{Lauberts \&
  Valentijn}{1989}]{LaubertsValentijn1989}
Lauberts A.,  Valentijn E.~A.,  1989, The surface photometry catalogue of the
  {ESO}-{Uppsala} galaxies.
\url {https://ui.adsabs.harvard.edu/abs/1989spce.book.....L}

\bibitem[\protect\citeauthoryear{Loubser \& Soechting}{Loubser \&
  Soechting}{2013}]{LoubserSoechting2013}
Loubser S.~I.,  Soechting I.~K.,  2013, \mn@doi [Monthly Notices of the Royal
  Astronomical Society] {10.1093/mnras/stt394}, 431, 2933

\bibitem[\protect\citeauthoryear{Lucy}{Lucy}{1974}]{Lucy1974}
Lucy L.~B.,  1974, \mn@doi [The Astronomical Journal] {10.1086/111605}, 79, 745

\bibitem[\protect\citeauthoryear{Luridiana, Morisset  \& Shaw}{Luridiana
  et~al.}{2015}]{LuridianaEtAl2015}
Luridiana V.,  Morisset C.,   Shaw R.~A.,  2015, \mn@doi [Astronomy \&
  Astrophysics] {10.1051/0004-6361/201323152}, 573, A42

\bibitem[\protect\citeauthoryear{Macchetto, Pastoriza, Caon, Sparks,
  Giavalisco, Bender  \& Capaccioli}{Macchetto
  et~al.}{1996}]{MacchettoEtAl1996}
Macchetto F.,  Pastoriza M.,  Caon N.,  Sparks W.~B.,  Giavalisco M.,  Bender
  R.,   Capaccioli M.,  1996, \mn@doi [Astronomy and Astrophysics Supplement
  Series] {10.1051/aas:1996307}, 120, 463

\bibitem[\protect\citeauthoryear{Machacek, O'Sullivan, Randall, Jones  \&
  Forman}{Machacek et~al.}{2010}]{MachacekEtAl2010}
Machacek M.~E.,  O'Sullivan E.,  Randall S.~W.,  Jones C.,   Forman W.~R.,
  2010, \mn@doi [The Astrophysical Journal] {10.1088/0004-637X/711/2/1316},
  711, 1316

\bibitem[\protect\citeauthoryear{Magorrian et~al.,}{Magorrian
  et~al.}{1998}]{MagorrianEtAl1998}
Magorrian J.,  et~al., 1998, \mn@doi [The Astronomical Journal]
  {10.1086/300353}, 115, 2285

\bibitem[\protect\citeauthoryear{Mauch, Murphy, Buttery, Curran, Hunstead,
  Piestrzynski, Robertson  \& Sadler}{Mauch et~al.}{2003}]{MauchEtAl2003}
Mauch T.,  Murphy T.,  Buttery H.~J.,  Curran J.,  Hunstead R.~W.,
  Piestrzynski B.,  Robertson J.~G.,   Sadler E.~M.,  2003, \mn@doi [Monthly
  Notices of the Royal Astronomical Society]
  {10.1046/j.1365-8711.2003.06605.x}, 342, 1117

\bibitem[\protect\citeauthoryear{Menezes, Ricci, Steiner, da Silva, Ferrari
  \& Borges}{Menezes et~al.}{2019}]{MenezesEtAl2019}
Menezes R.~B.,  Ricci T.~V.,  Steiner J.~E.,  da Silva P.,  Ferrari F.,
  Borges B.~W.,  2019, \mn@doi [Monthly Notices of the Royal Astronomical
  Society] {10.1093/mnras/sty3337}, 483, 3700

\bibitem[\protect\citeauthoryear{Menezes, Steiner, Ricci  \& da Silva}{Menezes
  et~al.}{2022}]{MenezesEtAl2022}
Menezes R.~B.,  Steiner J.~E.,  Ricci T.~V.,   da Silva P.,  2022, \mn@doi
  [Monthly Notices of the Royal Astronomical Society] {10.1093/mnras/stac1235},
  513, 5935

\bibitem[\protect\citeauthoryear{Nagar, Falcke  \& Wilson}{Nagar
  et~al.}{2005}]{NagarEtAl2005}
Nagar N.~M.,  Falcke H.,   Wilson A.~S.,  2005, \mn@doi [Astronomy \&
  Astrophysics] {10.1051/0004-6361:20042277}, 435, 521

\bibitem[\protect\citeauthoryear{Nayakshin \& Zubovas}{Nayakshin \&
  Zubovas}{2012}]{NayakshinZubovas2012}
Nayakshin S.,  Zubovas K.,  2012, \mn@doi [Monthly Notices of the Royal
  Astronomical Society] {10.1111/j.1365-2966.2012.21950.x}, 427, 372

\bibitem[\protect\citeauthoryear{Negus, Comerford, Sánchez, Revalski, Riffel,
  Bundy, Nevin  \& Rembold}{Negus et~al.}{2023}]{NegusEtAl2023}
Negus J.,  Comerford J.~M.,  Sánchez F.~M.,  Revalski M.,  Riffel R.~A.,
  Bundy K.,  Nevin R.,   Rembold S.~B.,  2023, \mn@doi [The Astrophysical
  Journal] {10.3847/1538-4357/acb772}, 945, 127

\bibitem[\protect\citeauthoryear{Osterbrock \& Ferland}{Osterbrock \&
  Ferland}{2006}]{OsterbrockFerland2006}
Osterbrock D.~E.,  Ferland G.~J.,  2006, Astrophysics of gaseous nebulae and
  active galactic nuclei, 2nd ed edn.
University Science Books, Sausalito, Calif

\bibitem[\protect\citeauthoryear{Papaderos et~al.,}{Papaderos
  et~al.}{2013}]{PapaderosEtAl2013}
Papaderos P.,  et~al., 2013, \mn@doi [Astronomy \& Astrophysics]
  {10.1051/0004-6361/201321681}, 555, L1

\bibitem[\protect\citeauthoryear{Pizzella et~al.,}{Pizzella
  et~al.}{1997}]{PizzellaEtAl1997}
Pizzella A.,  et~al., 1997, Astronomy and Astrophysics, v.323, p.349-356, 323,
  349

\bibitem[\protect\citeauthoryear{Ramella, Focardi  \& Geller}{Ramella
  et~al.}{1996}]{RamellaEtAl1996}
Ramella M.,  Focardi P.,   Geller M.~J.,  1996, Astronomy and Astrophysics,
  312, 745

\bibitem[\protect\citeauthoryear{Rampazzo, Panuzzo, Vega, Marino, Bressan  \&
  Clemens}{Rampazzo et~al.}{2013}]{RampazzoEtAl2013}
Rampazzo R.,  Panuzzo P.,  Vega O.,  Marino A.,  Bressan A.,   Clemens M.~S.,
  2013, \mn@doi [Monthly Notices of the Royal Astronomical Society]
  {10.1093/mnras/stt475}, 432, 374

\bibitem[\protect\citeauthoryear{Revalski, Crenshaw, Kraemer, Fischer, Schmitt
  \& Machuca}{Revalski et~al.}{2018}]{RevalskiEtAl2018}
Revalski M.,  Crenshaw D.~M.,  Kraemer S.~B.,  Fischer T.~C.,  Schmitt H.~R.,
  Machuca C.,  2018, \mn@doi [The Astrophysical Journal]
  {10.3847/1538-4357/aab107}, 856, 46

\bibitem[\protect\citeauthoryear{Ricci, Steiner  \& Menezes}{Ricci
  et~al.}{2014a}]{RicciEtAl2014}
Ricci T.~V.,  Steiner J.~E.,   Menezes R.~B.,  2014a, \mn@doi [Monthly Notices
  of the Royal Astronomical Society] {10.1093/mnras/stu441}, 440, 2419

\bibitem[\protect\citeauthoryear{Ricci, Steiner  \& Menezes}{Ricci
  et~al.}{2014b}]{RicciEtAl2014a}
Ricci T.~V.,  Steiner J.~E.,   Menezes R.~B.,  2014b, \mn@doi [Monthly Notices
  of the Royal Astronomical Society] {10.1093/mnras/stu442}, 440, 2442

\bibitem[\protect\citeauthoryear{Ricci, Steiner  \& Menezes}{Ricci
  et~al.}{2015a}]{RicciEtAl2015}
Ricci T.~V.,  Steiner J.~E.,   Menezes R.~B.,  2015a, \mn@doi [Monthly Notices
  of the Royal Astronomical Society] {10.1093/mnras/stu2576}, 447, 1504

\bibitem[\protect\citeauthoryear{Ricci, Steiner  \& Menezes}{Ricci
  et~al.}{2015b}]{RicciEtAl2015a}
Ricci T.~V.,  Steiner J.~E.,   Menezes R.~B.,  2015b, \mn@doi [Monthly Notices
  of the Royal Astronomical Society] {10.1093/mnras/stv1156}, 451, 3728

\bibitem[\protect\citeauthoryear{Ricci, Steiner, Menezes, Slodkowski Clerici
  \& da Silva}{Ricci et~al.}{2023}]{RicciEtAl2023}
Ricci T.~V.,  Steiner J.~E.,  Menezes R.~B.,  Slodkowski Clerici K.,
  da Silva M.~D.,  2023, \mn@doi [Monthly Notices of the Royal Astronomical
  Society] {10.1093/mnras/stad1130}, 522, 2207

\bibitem[\protect\citeauthoryear{Richardson}{Richardson}{1972}]{Richardson1972}
Richardson W.~H.,  1972, \mn@doi [JOSA] {10.1364/JOSA.62.000055}, 62, 55

\bibitem[\protect\citeauthoryear{Rickes, Pastoriza  \& Bonatto}{Rickes
  et~al.}{2008}]{RickesEtAl2008}
Rickes M.~G.,  Pastoriza M.~G.,   Bonatto C.,  2008, \mn@doi [Monthly Notices
  of the Royal Astronomical Society] {10.1111/j.1365-2966.2007.12724.x}, 384,
  1427

\bibitem[\protect\citeauthoryear{Riffel et~al.,}{Riffel
  et~al.}{2019}]{RiffelEtAl2019a}
Riffel R.~A.,  et~al., 2019, \mn@doi [Monthly Notices of the Royal Astronomical
  Society] {10.1093/mnras/stz841}, 485, 5590

\bibitem[\protect\citeauthoryear{Riffel et~al.,}{Riffel
  et~al.}{2021a}]{RiffelEtAl2021a}
Riffel R.~A.,  et~al., 2021a, \mn@doi [Monthly Notices of the Royal
  Astronomical Society] {10.1093/mnrasl/slaa194}, 501, L54

\bibitem[\protect\citeauthoryear{Riffel et~al.,}{Riffel
  et~al.}{2021b}]{RiffelEtAl2021}
Riffel R.,  et~al., 2021b, \mn@doi [Monthly Notices of the Royal Astronomical
  Society] {10.1093/mnras/staa3907}, 501, 4064

\bibitem[\protect\citeauthoryear{Riffel, Dors, Krabbe  \& Esteban}{Riffel
  et~al.}{2021c}]{RiffelEtAl2021b}
Riffel R.~A.,  Dors O.~L.,  Krabbe A.~C.,   Esteban C.,  2021c, \mn@doi
  [Monthly Notices of the Royal Astronomical Society] {10.1093/mnrasl/slab064},
  506, L11

\bibitem[\protect\citeauthoryear{Riffel, Riffel, Storchi-Bergmann, Costa-Souza,
  Souza-Oliveira  \& Bianchin}{Riffel et~al.}{2024}]{RiffelEtAl2024}
Riffel R.~A.,  Riffel R.,  Storchi-Bergmann T.,  Costa-Souza J.~H.,
  Souza-Oliveira G.~L.,   Bianchin M.,  2024, \mn@doi [Monthly Notices of the
  Royal Astronomical Society] {10.1093/mnras/stae055}, 528, 1476

\bibitem[\protect\citeauthoryear{Rodríguez del Pino, Arribas,
  Piqueras López, Villar-Martín  \& Colina}{Rodríguez del Pino
  et~al.}{2019}]{RodriguezdelPinoEtAl2019}
Rodríguez del Pino B.,  Arribas S.,  Piqueras López J.,  Villar-Martín
  M.,   Colina L.,  2019, \mn@doi [Monthly Notices of the Royal Astronomical
  Society] {10.1093/mnras/stz816}, 486, 344

\bibitem[\protect\citeauthoryear{Rose et~al.,}{Rose
  et~al.}{2019}]{RoseEtAl2019}
Rose T.,  et~al., 2019, \mn@doi [Monthly Notices of the Royal Astronomical
  Society] {10.1093/mnras/stz2138}, 489, 349

\bibitem[\protect\citeauthoryear{Rose et~al.,}{Rose
  et~al.}{2024}]{RoseEtAl2024}
Rose T.,  et~al., 2024, \mn@doi [Monthly Notices of the Royal Astronomical
  Society] {10.1093/mnras/stae1831}, 533, 771

\bibitem[\protect\citeauthoryear{Ruschel-Dutra \& Oliveira}{Ruschel-Dutra \&
  Oliveira}{2020}]{Ruschel-DutraOliveira2020}
Ruschel-Dutra D.,  Oliveira B. D.~D.,  2020, danielrd6/ifscube: {Modeling},
  \mn@doi{10.5281/ZENODO.4065550}, \url {https://zenodo.org/record/4065550}

\bibitem[\protect\citeauthoryear{Ruschel-Dutra et~al.,}{Ruschel-Dutra
  et~al.}{2021}]{Ruschel-DutraEtAl2021}
Ruschel-Dutra D.,  et~al., 2021, \mn@doi [Monthly Notices of the Royal
  Astronomical Society] {10.1093/mnras/stab2058}, 507, 74

\bibitem[\protect\citeauthoryear{Sarzi et~al.,}{Sarzi
  et~al.}{2010}]{SarziEtAl2010}
Sarzi M.,  et~al., 2010, \mn@doi [Monthly Notices of the Royal Astronomical
  Society] {10.1111/j.1365-2966.2009.16039.x}, 402, 2187

\bibitem[\protect\citeauthoryear{Schlafly \& Finkbeiner}{Schlafly \&
  Finkbeiner}{2011}]{SchlaflyFinkbeiner2011}
Schlafly E.~F.,  Finkbeiner D.~P.,  2011, \mn@doi [The Astrophysical Journal]
  {10.1088/0004-637X/737/2/103}, 737, 103

\bibitem[\protect\citeauthoryear{Segers, Schaye, Bower, Crain, Schaller  \&
  Theuns}{Segers et~al.}{2016}]{SegersEtAl2016}
Segers M.~C.,  Schaye J.,  Bower R.~G.,  Crain R.~A.,  Schaller M.,   Theuns
  T.,  2016, \mn@doi [Monthly Notices of the Royal Astronomical Society:
  Letters] {10.1093/mnrasl/slw111}, 461, L102

\bibitem[\protect\citeauthoryear{She, Ho  \& Feng}{She
  et~al.}{2017}]{SheEtAl2017}
She R.,  Ho L.~C.,   Feng H.,  2017, \mn@doi [The Astrophysical Journal]
  {10.3847/1538-4357/835/2/223}, 835, 223

\bibitem[\protect\citeauthoryear{Singh et~al.,}{Singh
  et~al.}{2013}]{SinghEtAl2013}
Singh R.,  et~al., 2013, \mn@doi [Astronomy \& Astrophysics]
  {10.1051/0004-6361/201322062}, 558, A43

\bibitem[\protect\citeauthoryear{Slee, Sadler, Reynolds  \& Ekers}{Slee
  et~al.}{1994}]{SleeEtAl1994}
Slee O.~B.,  Sadler E.~M.,  Reynolds J.~E.,   Ekers R.~D.,  1994, \mn@doi
  [Monthly Notices of the Royal Astronomical Society]
  {10.1093/mnras/269.4.928}, 269, 928

\bibitem[\protect\citeauthoryear{Stasińska et~al.,}{Stasińska
  et~al.}{2008}]{StasinskaEtAl2008}
Stasińska G.,  et~al., 2008, \mn@doi [Monthly Notices of the Royal
  Astronomical Society: Letters] {10.1111/j.1745-3933.2008.00550.x}, 391, L29

\bibitem[\protect\citeauthoryear{Steiner, Menezes, Ricci  \& Oliveira}{Steiner
  et~al.}{2009}]{SteinerEtAl2009}
Steiner J.~E.,  Menezes R.~B.,  Ricci T.~V.,   Oliveira A.~S.,  2009, \mn@doi
  [Monthly Notices of the Royal Astronomical Society]
  {10.1111/j.1365-2966.2009.14530.x}, 395, 64

\bibitem[\protect\citeauthoryear{Steiner et~al.,}{Steiner
  et~al.}{2022}]{SteinerEtAl2022}
Steiner J.~E.,  et~al., 2022, \mn@doi [Monthly Notices of the Royal
  Astronomical Society] {10.1093/mnras/stac034}, 510, 5780

\bibitem[\protect\citeauthoryear{Tody}{Tody}{1986}]{Tody1986}
Tody D.,  1986, in Crawford D.~L.,  ed.,  Society of {Photo}-{Optical}
  {Instrumentation} {Engineers} ({SPIE}) {Conference} {Series} Vol. 627,
  Instrumentation in astronomy {VI}. Crawford, David L., Tucson, p.~733,
  \mn@doi{10.1117/12.968154}, \url
  {http://proceedings.spiedigitallibrary.org/proceeding.aspx?doi=10.1117/12.968154}

\bibitem[\protect\citeauthoryear{Tody}{Tody}{1993}]{Tody1993}
Tody D.,  1993, in Astronomical {Data} {Analysis} {Software} and {Systems} {I}.
  p.~173, \url {https://ui.adsabs.harvard.edu/abs/1993ASPC...52..173T}

\bibitem[\protect\citeauthoryear{Tully et~al.,}{Tully
  et~al.}{2013}]{TullyEtAl2013}
Tully R.~B.,  et~al., 2013, \mn@doi [The Astronomical Journal]
  {10.1088/0004-6256/146/4/86}, 146, 86

\bibitem[\protect\citeauthoryear{Veilleux \& Osterbrock}{Veilleux \&
  Osterbrock}{1987}]{VeilleuxOsterbrock1987}
Veilleux S.,  Osterbrock D.~E.,  1987, \mn@doi [The Astrophysical Journal
  Supplement Series] {10.1086/191166}, 63, 295

\bibitem[\protect\citeauthoryear{Veron-Cetty \& Veron}{Veron-Cetty \&
  Veron}{1988}]{Veron-CettyVeron1988}
Veron-Cetty M.-P.,  Veron P.,  1988, Astronomy and Astrophysics, Vol. 204, p.
  28-38 (1988), 204, 28

\bibitem[\protect\citeauthoryear{Yan \& Blanton}{Yan \&
  Blanton}{2012}]{YanBlanton2012}
Yan R.,  Blanton M.~R.,  2012, \mn@doi [The Astrophysical Journal]
  {10.1088/0004-637X/747/1/61}, 747, 61

\bibitem[\protect\citeauthoryear{Zeilinger et~al.,}{Zeilinger
  et~al.}{1996}]{ZeilingerEtAl1996}
Zeilinger W.~W.,  et~al., 1996, \mn@doi [Astronomy and Astrophysics Supplement
  Series] {10.1051/aas:1996290}, 120, 257

\bibitem[\protect\citeauthoryear{de Vaucouleurs, de Vaucouleurs, Corwin, Buta,
  Paturel  \& Fouque}{de~Vaucouleurs et~al.}{1991}]{deVaucouleursEtAl1991}
de Vaucouleurs G.,  de Vaucouleurs A.,  Corwin Herold~G. J.,  Buta R.~J.,
  Paturel G.,   Fouque P.,  1991, Third reference catalogue of bright galaxies.
Springer-Verlag, New York

\bibitem[\protect\citeauthoryear{do Nascimento et~al.,}{do~Nascimento
  et~al.}{2019}]{doNascimentoEtAl2019}
do Nascimento J.~C.,  et~al., 2019, \mn@doi [Monthly Notices of the Royal
  Astronomical Society] {10.1093/mnras/stz1083}, 486, 5075

\bibitem[\protect\citeauthoryear{van Dokkum}{van Dokkum}{2001}]{vanDokkum2001}
van Dokkum P.,  2001, \mn@doi [Publications of the Astronomical Society of the
  Pacific] {10.1086/323894}, 113, 1420

\makeatother
\end{thebibliography}








\bsp	
\label{lastpage}
\end{document}